# Immunity to Scaling in $MoS_2$ Transistors Using Edge Contacts

Zhihui Cheng[1], Katherine Price[1], Shreya Singh[1], Steven Noyce[1], Yuh-Chen Lin[1], Yifei Yu[2], Linyou Cao[2], Aaron D. Franklin[1,3]

[1]Department of Electrical and Computer Engineering, Duke University, Durham, NC 27708

[2]Department of Materials Science and Engineering, North Carolina State University, Raleigh, NC 27695

[3]Department of Chemistry, Duke University, Durham, NC 27708

**Abstract**

**Atomically thin two-dimensional (2D) materials are promising candidates for sub-10 nm transistor channels due to their ultrathin body thickness, which results in strong electrostatic gate control. Properly scaling a transistor technology requires reducing both the channel length (distance from source to drain) and the contact length (distance that source and drain interface with semiconducting channel). Contact length scaling remains an unresolved epidemic for transistor scaling, affecting devices from all semiconductors — silicon to 2D materials. Here, we show that clean edge contacts to 2D $MoS_2$ provide immunity to the contact-scaling problem, with performance that is independent of contact**

**length down to the 20 nm regime. Using a directional ion beam, *in situ* edge contacts of various metal-$MoS_2$ interfaces are studied. Characterization of the intricate edge interface using cross-sectional electron microscopy reveals distinct morphological effects on the $MoS_2$ depending on its thickness — from monolayer to few-layer films. Chromium is found to outperform other metals in the edge contact scheme, which is attributed to the shorter Cr-$MoS_2$ bond length. Compared to scaled top contacts with 20 nm contact length, *in situ* edge contacts yield better performance with an effective contact length of ~ 1 nm and 18 times higher carrier injection efficiency. The *in situ* edge contacts also exhibit ~8 times higher performance compared to the best-reported edge contacts. Our work provides experimental evidence for a solution to contact scaling in transistors, using 2D materials with clean edge contact interfaces, opening a new way of designing devices with 2D materials.**

Booming applications, such as smartphones, autonomous vehicles, and server farms, leave society starving for more computational power. At the heart of virtually all computation is the transistor, which yields increased computational ability with each successive technology node through size scaling. Such scaling, which enjoyed decades of success predicted by Moore's law, is now undisputedly slowing and potentially reaching an end based on the limitations of silicon[1–5]. Not surprisingly, the electronic device community has been eager to explore new materials for the transistor channel that may extend the scalability roadmap, even for a few more generations. Nanomaterials have long been seen as a viable option, from 1D carbon nanotubes to the expanding family of 2D crystals. For 2D, graphene initially captured widespread attention and spawned a whole library of 2D materials with a variety of electronic band structures and properties[6–9].

The main advantage of 2D materials is their ultra-thin nature, which could enable extremely scaled transistors for the "Beyond Moore" era. The ultrathin body thickness directly affects the screening length, which dictates how short the channel length can be scaled down without inducing deleterious short channel effects. Using a planar device structure, it is estimated that monolayer $MoS_2$ has a screening length of less than 1 nm[10], assuming an equivalent oxide thickness (EOT) of 1 nm is used. This suggests that the gate-tunable, 2D-based transistor can be scaled to sub-5 nm channel length — a scale where Si encounters severe short channel effects using similar gate structures. Both experimental and theoretical studies have demonstrated the superb channel length scalability of 2D field-effect transistors (FETs)[10–16]. Aside from the superior scalability, 2D materials also offer new possibilities for other unconventional applications (for example, flexible electronics) because of their substrate independence[17–19]. Moreover, a plethora of atomic heterostructures can be formed between different 2D materials[20–24], in a way that is inaccessible to traditional semiconductors.

While the channel length scalability of 2D FETs has been well studied[10–15], the contact length scaling and its related challenges have been largely neglected. However, contact engineering in general for 2D FETs has been a topic of considerable interest, including using different metals[25,26], transforming phases[27], and controlling metal deposition conditions[28]. While these approaches deepen our understanding of the metal-2D interface and have achieved contact resistance as low as 200 Ω•µm, they all use a contact length of at least hundreds of nanometers, which are orders of magnitude larger than needed for actual technologies. A fully scaled device technology for the 2030 era will need both the channel and contact lengths scaled below 12 nm (equivalent to a contacted gate pitch of 24 nm)[29]. Note that contact scaling is an epidemic for all semiconductors, including Si. In a Si FinFET, two thirds of the gate pitch (54 nm for Intel's 10 nm node technology) is the contact length (36 nm)[30]. Since future scaled transistors would have a shorter gate pitch, the shrinking gate pitch also leads to shrinking contact length, thus decreasing on-state performance[31] and highlighting the importance of improving contact scaling behavior for all types of transistors. In a simplified top-contacted and back-gated $MoS_2$ transistor, as shown in Fig. 1a, as the contact length ($L_c$) decreases, the area available for carrier injection is also reduced. The shrinking contact length leads to severely degraded performance, especially when $L_c$ drops below the transfer length ($L_T$ = 30 ~ 40 nm for $MoS_2$[28], as depicted in Fig.1b), which is the length over which the majority of carriers are injected.

Ideally, for scaling, contacts would be bonded directly to the side of the 2D channel as pure "edge contacts," as illustrated in Fig.1c, where charge is injected from the metal directly into the 2D crystal via covalent bonds. Since the area of injection at the edge is independent of the physical contact length, we hypothesize that edge contact could provide ultimate scalability, as shown hypothetically in Fig. 1d, where the on-current ($I_d$) would be independent of the $L_c$. Several studies on edge contacts to 2D materials have been reported, beginning with Cr edge contacts to graphene that exhibited a low contact resistance of 150 Ω•μm[32], though graphene is not a semiconductor. In a separate study[33], an edge-like interface between graphene and $MoS_2$ was demonstrated; however, the scalability of the 2D-2D hetero-junction remained uncertain as the $L_c$ demonstrated is over 20 μm. Moreover, growing the graphene-$MoS_2$ edge added additional complexity and variability to the fabrication process, reducing the reliability of this approach. Finally, demonstration of edge contacts between metal and $MoS_2$ has been limited to the use of an *ex situ* and isotropic plasma etching approach[34]. The performance metrics such as on-current and on-off ratio were unfavorable, possibly due to the uncleanliness of the interface with the dangling bonds in the exposed $MoS_2$ edge reacting with species in the ambient owing to the use of an *ex situ* plasma etching. Considering the ultra-sensitive nature of the dangling bonds at the edge, it is thus crucial to have the interface preserved in a clean *in situ* environment in order to properly determine the potential of metal-$MoS_2$ edge contacts.

Here, we demonstrate edge-contacted $MoS_2$ FETs by using an *in situ* Ar ion beam. We show the ultimate scalability of pure edge contacts to CVD-grown $MoS_2$ of various layer thicknesses and metal types, providing evidence for the immunity of edge contacted 2D FETs to aggressive contact scaling. In order to understand carrier transport, we use cross-sectional scanning transmission electron microscopy (STEM) and low-temperature electrical measurement to characterize the edge contacts. Our study elucidates the intriguing metal-2D edge interface and the potential of edge contacts for future scaled transistors.

**Etching Capability of a Directional Argon Ion Beam**

The use of an *in situ* ion beam to etch the $MoS_2$ immediately prior to contact metallization is crucial to avoid reactivity between the created edge states and molecular species other than the contact metal. The *in situ* ion beam source is incorporated with an electron beam evaporator in the same ultra high vacuum (UHV) chamber, as shown in Fig. 2a. The etching effect of the ion beam on $MoS_2$ is studied using the process shown in Fig. 2b. Selective bombardment of the exposed (contact) regions by the directional Ar ion beam is achieved using patterned PMMA (which shields the channel regions). Note that the Ar ion beam has a minimal etching effect on the PMMA, which make PMMA a suitable etch mask, as shown in Supplementary Fig. 1. Our previous study[35] shows that low-energy (~100 eV) Ar ion bombardment can create vacancies in the 2D crystal. Here, a higher energy (~600 eV) ion beam is shown to controllably etch the $MoS_2$, as shown in the atomic force microscopy (AFM) image in Fig. 2c. We also use energy dispersive spectroscopy (EDS) to map the etched flake in Fig. 2c. The sulfur signal in Fig. 2d and molybdenum signal in Fig. 2e further prove the etching capability of the Ar ion beam. The AFM profiles and line scans from different regions show how both the $MoS_2$ and $SiO_2$ are etched by the ion bombardment, as plotted in Fig. 2f. Note that the edge of $MoS_2$ in the etched region attracts more reacted species/residue (as high as 100 nm in Fig. 2g), evidential of the higher reactivity of the $MoS_2$ edge when exposed to solvent/air (*ex situ*) and the importance of forming edge contacts with an *in situ* process. Meanwhile, the flake edge that has not been exposed to the ion beam is relatively clean, as shown in Fig. 2g. This further exemplifies the highly reactive etched edge, which could be useful in other applications such as sensing since it could act as a preferable binding site for antibodies compared to either the basal surface that has limited dangling bonds or the natural edges that are less reactive. In Fig. 2g, we also label the $SiO_2$ and $MoS_2$ etched depth shown in Fig. 2f. The linear relationship between the etch-depth and the ion beam exposure time is plotted in Supplementary Fig. 2, showing an etching rate of 1.83 Å/s for $MoS_2$ and 1.2 Å/s for $SiO_2$.

**Edge Contacts to exfoliated multilayer $MoS_2$**

Upon exposing the $MoS_2$ edge in the contact regions under UHV, contact metal is then deposited using an electron beam evaporator in the same chamber. The newly generated edge states are able to react with the depositing metal, forming a bonded edge interface. To study this interface, we use cross-sectional scanning tunneling electron microscopy (STEM) to characterize the etched edge. Fifteen layers of $MoS_2$ were exfoliated onto a silicon wafer with 300 nm $SiO_2$ (see cross-sectional STEM image in Supplementary Fig. 3). After using the etching process illustrated in the last section, the metal contact was *in situ* deposited on the etched region (Fig. 3a). The cross-sectional STEM image of the finished contact is shown in Fig. 3b. The etching process creates the unique splitting and tapering effects (Fig. 3c), which is particularly surprising as these effects are different from the common undercut[36] and microtrench[37] profile seen in some isotropic *ex situ* plasma processes. The splitting effect could be attributed to the interaction between the directional Ar ion beam and the weak van der Waals interlayer binding of the 2D materials. The splitting effect could profoundly change electronic properties of the $MoS_2$ at the edge (further details in Supplementary Note 1). Meanwhile, the tapering effect is common for directional dry etching[38], as the center region receives more directional ion bombardment. These effects open a new window of opportunities to study the intricate interface between metal and 2D materials and to use in other applications such as sensing and material intercalation[39–41].

To further understand the metal-$MoS_2$ edge interface, EDS was used to characterize the elements present in the right-side edge of the contact. As shown in Figure 3d, the $MoS_2$ is topped with 2 nm of Ti (green) and 20 nm of Au (red). The thickness of Ti is more uniform in the area where there is more $MoS_2$ edge in the splitting and tapering region, indicative of more consistent bonding because of the reactive edge states. Additionally, we noticed the presence of sulfur (turquoise) at the junction of the metal-$MoS_2$ edge in this splitting region, where crystalline $MoS_2$ has already ended (Fig. 3d). These sulfur-metal hybrid areas could indicate the covalent bond between Ti/Au and sulfur. Also, the oxygen element was mapped in Supplementary Fig. 4 and no higher concentration of oxygen appears in the interface between Ti and $MoS_2$, which suggests that the *in situ* environment is relatively pristine.

In addition to the interface highlighted in Fig. 3, where the full multilayer $MoS_2$ is etched by the Ar ion beam in the center of the contact regions (quasi-edge contacts), we also used shorter etching time (25 and 50 s) to produce partially etched $MoS_2$ in the center of the contact region (partial-edge contacts), as given in Supplementary Fig. 5. Since the exfoliated flake is about 10 nm thick (15L), the tapering and splitting effects in Fig. 3 also show up in the partial-edge contacts. We then fabricated devices on multilayer flakes with different thickness (35 and 8 nm) in order to compare performance of the quasi-edge and partial-edge contacts (see Supplementary Note 1-2). Compared to the partial-edge contacts (9 μA/μm at $V_{ds}$=1 V), quasi-edge contacts yield smaller current (5 μA/μm at $V_{ds}$=1 V) but have a distinct forming or "burn-in" effect when large $V_{ds}$ (over 3 V) is applied. This forming behavior suggests that a large electric field from source to drain can strength the bond between the metal and $MoS_2$ edge states. Considering that the defects created on the tapering region add additional complications to the analysis, further investigation is needed to resolve the carrier injection through the splitting $MoS_2$ edge and the tapering layers. In the following section, in order to demonstrate pure edge contacts and their scaling behavior, we focus on CVD-grown $MoS_2$ films since they offer a large area of thin crystals (1-4 layers with size of over 100 μm$^2$).

**Edge Contacts to CVD-Grown $MoS_2$**

In order to demonstrate the ultimate scalability of edge contacts, *in situ* edge contacts were fabricated on CVD-grown $MoS_2$. These $MoS_2$ films have a large area with uniform thickness, making them suitable for device fabrication and performance comparison. Trilayer and monolayer CVD films were used to fabricate *in situ* edge contacts as shown in Fig. 4. These films were grown directly onto $SiO_2$ without the need of a transfer process, which could introduce contaminants such as water molecules and resist residue. In Fig. 4a, a small rectangular box of $MoS_2$ was used, as the materials outside of the rectangular box are etched away using $CF_4$ plasma. After an e-beam lithography process, the same Ar ion beam etching process to Fig. 2b with an etching time of 30 s was used and the contact metal (Ni) was deposited *in situ* inside the same UHV chamber. A diagram of scaled edge contacts to $MoS_2$ is given in Fig. 4b, where two long contacts ($L_c$ = 60 nm) and two short contacts ($L_c$ = 20 nm) were fabricated onto the same film. The cross-sectional STEM image of the right-side of the $L_c$ = 60 nm edge contacts is shown in Fig. 4c. The metal entrenches into the oxide and contacts the edge of the trilayer film without the splitting effect, producing pure edge contacts. The side-view of the 3-layer $MoS_2$ film with atomic resolution is given in Fig. 4d, showing the crystal structure of the 2D material. Characterization of the devices with different contact lengths (Fig. 4e,f) revealed that the $L_c$ = 20 nm and $L_c$ = 60 nm FETs have the essentially same $I_d$, independent of the contact length. These edge-contacted trilayer devices outperform their top-contacted trilayer device counterparts (all device dimensions and materials being the same), with $I_{on}$ = 10 µA/µm at overdrive voltage $V_{ov} = V_{gs} - V_{th}$ = 30 V and $V_{ds}$ = 4 V (see Supplementary Fig. 6). One of the most encouraging aspects of this result is the sheer density of carriers being injected into the edge contact area (effective $L_c$ = 1 nm), which is over an order of magnitude smaller than the top contact $L_c$ using the same film and two orders of magnitude smaller than the top contact $L_c$ used in other studies. Note that when comparing results with different studies, all of the relevant variables need to be considered, such as the film quality, film thickness, oxide type, oxide thickness, metal evaporation conditions, overdrive voltage and the drain voltage $V_{ds}$, at which the current is extracted. The high variability for devices built on $SiO_2$-grown $MoS_2$ films (see Supplementary Table 1) also needs to be considered[42]. For example, even for exfoliated monolayer $MoS_2$, the contact resistance can range from several kΩ•µm to 100 kΩ•µm[28]. Because our top and edge contacted devices are built using the same conditions, our results represent the potential of ultimate contact scaling using *in situ* edge contacts.

Edge contacts to monolayer $MoS_2$ from CVD-grown crystals were also explored. A device structure similar to the one illustrated in Fig. 4a was used, with monolayer $MoS_2$ as the channel material (Fig. 4g). A triangular monolayer film was chosen and the same process of *in situ* etching and metal evaporation was used to make the edge contacts with different contact lengths.

The cross-sectional STEM images show the metal entrenching into the oxide, representative of complete $MoS_2$ removal in the contact region. EDS images of the contact (Fig. 4i) provide further evidence of the isolation of the $MoS_2$ to the channel and the abrupt contact interface. A magnified view of the sulfur at the edge is given in Supplementary Fig. 7, further showing this abrupt cut-off of the monolayer $MoS_2$ at the edge. The corresponding $I_d$ – $V_{gs}$ curves for the monolayer $MoS_2$ devices are given in Fig. 4k-l.

We also compare the $I_d$ – $V_{ds}$ characteristics for the monolayer edge and top contacted devices in Supplementary Fig. 8. The performance of these edge-contacted monolayer devices is within the same range as their top-contacted counterparts using the same metal and $MoS_2$ film. Compared to trilayer $MoS_2$, monolayer devices (both top- and edge-contacted) suffer greatly from the interface traps formed between $MoS_2$ and $SiO_2$ in the high temperature growth process (750 °C). $MoS_2$ films grown on other substrates (for example, sapphire) and then transferred to $SiO_2$ substrates could offer less variability and higher performance (see Supplementary Fig. 9).

**Ultimate Contact Scaling**

A scaling comparison between top and edge contacts is essential to determine the advantages of the edge contact scheme. On the multilayer CVD-grown $MoS_2$ flakes, Cr top contacts and *in situ* Cr edge contacts were fabricated. The performance comparison between scaled Cr top and edge contacts is shown in Figure 5(a-b), where the on-state performance of edge contacts (both $L_c$ = 20 nm and 60 nm) is ~18 μA/μm, at $V_{ov}$ = 30 V and $V_{ds}$ = 4 V. The consistency in the performance of the $L_c$ = 60 nm edge-contacted device with that of the 20 nm one is indicative of the true edge profile and pure edge injection of carriers. Even though $I_d$ of $L_c$ = 20 nm Cr top contacts is similar to the performance of the edge contacts at large $V_{ds}$ = 4V, attention should be given to the device performance at a low $V_{ds}$, where we can learn more information on the carrier injection behavior in the contacts. In Fig. 5(c), we plotted the $I_d$ versus $L_c$ at $V_{ds}$ = 0.5 V and $V_{ov}$ = 30 V. The total resistance $R_{tot}$ was placed on the right axis, showing an inverse relationship with the $I_d$ on the left axis. The fitting curves for top Ni and Cr do not saturate within the $L_c$ < 100 nm range, which is explained in the Supplementary Note 4. Not surprisingly, top Cr contacts with short contact length ($L_c$ = 20 nm) have a much higher $R_{tot}$ than the top contacts with long contact length ($L_c$ = 60 nm). This trend is also true for $R_c$ ($R_c = (R_{tot} - R_{ch})/2$) since the same $L_{ch}$ was used for all devices in Fig. 5(c) and the resistance of the channel $R_{ch}$ relies on the $L_{ch}$ (for normalized contact width, $R_{ch} = R_{sh}L_{ch}$, with $R_{sh}$ being the sheet resistance of $MoS_2$ in the channel). This deterioration of $R_c$ on top Cr contacts presents the challenge of using top contacts for scaled devices. However, *in situ* edge contacts with different $L_c$ show relatively constant $R_{tot}$ because the carriers are injected through the edge, which is independent of $L_c$. When $L_c$ is at scaled dimension (< 20 nm), edge contacts demonstrate clear advantages over top contacts, for both Cr and Ni, providing immunity for the contact scaling. We also compare the *in situ* edge contacts with other reported *ex situ* edge contacts (see Supplementary Table 1). The reported $R_{tot}$ of the *ex situ* metal-$MoS_2$ edge contacts varies from 3 to hundreds of MΩ• μm. The current at $V_{ds}$ = 0.5 V of *in situ* edge contacts in this work is 7.8 times higher than the best-reported *ex situ* counterparts. Since $R_{tot} = 0.5\ V/ I_d$, $R_{tot}$ of the *in situ* Cr-$MoS_2$ edge contacts (500 KΩ• μm) is only 11.4% of the best-reported *ex situ* edge contacted devices. Even though the thickness of $MoS_2$ (3L for *in situ* Cr edge-contacted devices) is slightly thicker (1L for *ex situ* Sc edge counterparts), the *in situ* Cr-edge contacted devices only have about half of the carrier density in the *ex situ* Sc edge contacts ($2.16\times10^{12}$ cm$^{-2}$ versus $4.16\times10^{12}$ cm$^{-2}$). This improvement could be associated with the different metal types, the directional ion beam etching and *in situ* metal deposition. Compared with the best-reported edge contacts, the *in situ* Cr edge contacts demonstrate significant advances for better edge contacts to semiconducting 2D materials.

Estimating the $R_c$ of *in situ* edge contacts is also important. We first extracted $R_{sh}$ and $R_c$ for top Cr contacted devices from using transfer length model (TLM) structures (see Supplementary Fig. 10). The $R_c$ for the top contacts with $L_c$ = 60 nm is ~110 KΩ•μm and $R_{sh}$ is ~100 KΩ/square. Using this $R_{sh}$ for 3L CVD-grown $MoS_2$, we can estimate the $R_c$ for Cr edge contacts to 3L $MoS_2$ to be 205 KΩ•μm, which outperforms the $R_c$ for top Cr contacts with $L_c$ = 20 nm (381 KΩ•μm), as shown in Supplementary Table 2. It should be noted that the area for carrier injection ($A_{inj}$) in the scaled top contacts ($L_c$ = 20 nm) is 10 times larger than $A_{inj}$ for edge contacts (2 nm thick for 3L $MoS_2$). Combining the area of carrier injection and the contact resistance, carrier injection efficiency can be defined as $1/(A_{inj}•R_c)$, with the efficiency for edge contacts is at least 18 times higher than the scaled top contacts ($L_c$ = 20 nm). While the $R_c$ for edge and top contacts has been compared, we stress that these two $R_c$ are intrinsically different. The top contact resistance includes the interfacial resistance of the metal-$MoS_2$ interface ($\rho_c$) and the series resistance of the $MoS_2$ underneath the contact metal ($R_{sh}$), which resists lateral carrier flow beneath the metal contacts. For simplicity, we use the same label $R_{sh}$ for the sheet resistance in the channel and underneath the contact, assuming their values are close. In contrast, the edge contact resistance is solely the metal-$MoS_2$ edge resistance (see more details in Supplementary Note 4). These intrinsic differences merit further investigations using contact engineering approaches that maybe different from those developed for top contacts.

To further understand the *in situ* edge contact, we characterized Cr edge contacts under low-temperatures. As given in Supplementary Note 5, a Schottky barrier of 120 meV is extracted. The Arrhenious plot in Supplementary Notes Fig. 4 looks surprising. At high temperatures (300 to 250 K on the left side), the fitting curves are dropping, which is the evidence for thermionic transport over the Schottky barrier. But at low temperatures (below 200 K on the right side), regardless of the $V_{gs}$, the fitting curves go up, which suggests that carriers are tunneling through a barrier. The behavior of the curves at low temperature is abnormal because from the large amount of low-temperature characterizations on top contacts reported elsewhere, the fitting curves all turn downward at low $V_{gs}$ while only going flat or up when the $V_{gs}$ is large enough that Schottky barrier becomes thin and tunneling becomes dominant (see Supplementary Notes Fig. 4-5 for comparison). This unique Arrhenius profile suggests there is an additional tunneling route formed at the edge contacts that are independent of $V_{gs}$. A more focused, detailed analysis in subsequent studies is deserved to investigate the formation of this tunneling route and the impact of edge interfaces on the band diagram of the edge contacts.

The effect of different metal types is also important in understanding the *in situ* edge contact scheme. The *I-V* characteristics of Au, Cr, and Ni are compared in Supplementary Fig. 11. Cr outperforms the other metals, as similarly observed with edge contacts to graphene[32]. Theoretically, Cr has been proposed to be an ideal metal to contact $MoS_2$ in the top contact scheme, with its shorter bond length to S, larger binding energy, and larger density of state at $E_F$[45]. As the bonding length could be shorter in the edge contact scheme, density-functional theory (DFT) calculations on Cr edge contacts to $MoS_2$ remain to be conducted in order to confirm the orbital overlapping profile. Experimentally, devices with different metals have different threshold voltages, which can be explained by the different height and shape of the Schottky barrier for different metal-$MoS_2$ interfaces. We also compare the contact resistance for different metals in Supplementary Note 3. The huge contact resistance of Au edge contact compared to Au top contacts contradicts the suggestion that there might be some top interface transport component in the edge interface, otherwise the Au edge contact should perform similar to the Au top contacted devices.

Overall, while the top contacts can outperform *in situ* edge contacts at long contact lengths of $L_c$ > 20 nm, attention should be given to the short contact length where the 2D materials would most likely be utilized in future scaled transistors. Furthermore, now that edge contacts to a 2D semiconductor have been demonstrated, continued study and optimization will improve their quality and resulting device performance. Further investigations may include: 1) improving the film quality of the 2D materials to have fewer defects and higher mobility; 2) doping the contact region before fabricating the edge contacts to further increase the number of carriers injected to the flake through the edge and thus decrease the contact resistance[46]; and 3) exploring more metal types to find a preferable edge interface.

**Conclusion**

*In situ* edge contacts to $MoS_2$ FETs were demonstrated to provide immunity to contact length scaling for future generation devices. The challenge of preserving and utilizing the exposed, reactive edge of the $MoS_2$ was overcome by using *in situ* ion beam etching with contact metal deposition. The performance of the transistors remained consistent even as $L_c$ ranged from 20 nm to 60 nm across a set of devices, experimentally demonstrating that edge contacts are advantageous for ultimate 2D contact scaling. Moreover, the comparison of edge contacts versus top contacts was demonstrated and the impact of different metals (Ni, Cr, and Au) was explored using the same edge contact scheme. Further theoretical and experimental investigations are warranted to better understand the edge contact interface and decrease the contact resistance. Our work sheds light on the potential of edge contacts for ultimate contact scaling in $MoS_2$ transistors and could be applied to other 2D materials and nanoelectronic devices, paving the road for future aggressively scaled devices.

## Methods

**Growth of the $MoS_2$ by CVD.** The $MoS_2$ flakes were grown using a chemical vapor deposition (CVD) process reported previously[47–49]. Typically, 1g sulfur powder (Sigma-Aldrich) and 15-30mg $MoO_3$ (99.99%, Sigma-Aldrich) source material were placed upstream and at the center of a tube furnace, respectively. The substrates (heavily-doped Si substrate with 300 nm $SiO_2$) were placed downstream in the furnace tube. Typical growth was performed at 750 °C for 10 minutes under a flow of Ar gas in rate of 100 sccm and ambient pressure.

**Fabrication of *in situ* edge-contacted devices**

For devices using exfoliated flakes, multilayer $MoS_2$ flakes were mechanically exfoliated onto a heavily-doped Si substrate with 300 nm $SiO_2$. For devices using CVD-grown $MoS_2$ films, the $MoS_2$ crystal was grown using the above process. EBL with PMMA was used to define the contact regions, leads, and pads. The substrate was then developed in a solution of IPA:MIBK= 3:1. After developing, the substrate was transferred to the UHV chamber (base pressure ~$10^{-8}$ torr) having an ion beam source (KDC 40, KRI) *in situ* with an e-beam evaporator. The chip was exposed with a 600 eV directional Ar ion beam, followed by metal deposition. A top Au layer (30 nm) is also *in situ* deposited on top of the *in situ* Ni and Cr metal (normally 15 nm) to prevent oxidation of the contacts when exposed to ambient. This *in situ* ion beam process with metal deposition is crucial for protecting the exposed edges from other molecules in the ambient environment. Finally, the fabricated devices were characterized in ambient air after lift-off in acetone at a temperature of 80 °C.

**Characterization of the edge contact interface.** The AFM images in Fig. 2 were taken from a Digital Instruments Dimension 3100. The SEM images are obtained using an FEI XL30 SEM-FEG. The EDS images in Fig. 3 are from a Bruker XFlash 4010 EDS. The cross-sectional STEM images in Fig. 3 and 4 were obtained using the FEI Titan 80-300 probe aberration corrected STEM with monochromator. The EDS images in Fig. 3 and 4 were acquired from the SuperX system with the four Bruker Silicon Drift Detectors (SDD).

**Data Availability**. The data that support this work are within this paper and other findings of this study are available from the corresponding author upon reasonable request.

## Acknowledgments

This work is supported by National Science Foundation (NSF) under Grant ECCS 1508573. The authors would like to thank the staffs in Shared Instrument and Manufacturing Facilities (SMIF) at Duke for their assistance as well as Rohan Dhall in obtaining the STEM images in Analytical Instrumentation Facility (AIF) at NCSU.

## Author Contributions

A.F. conceived the idea and led the research. Z.C. design the experiment, fabricated and characterized the devices. K.P. and S.N. helped on designing the experiment. S.S helped on characterizing the devices. Y.Y. grow the $MoS_2$ using CVD. Z.C. and A.F. wrote the main paper and the Supplementary Information with input from all other authors.

## Competing Interests

The authors declare no competing interests.

## Figures

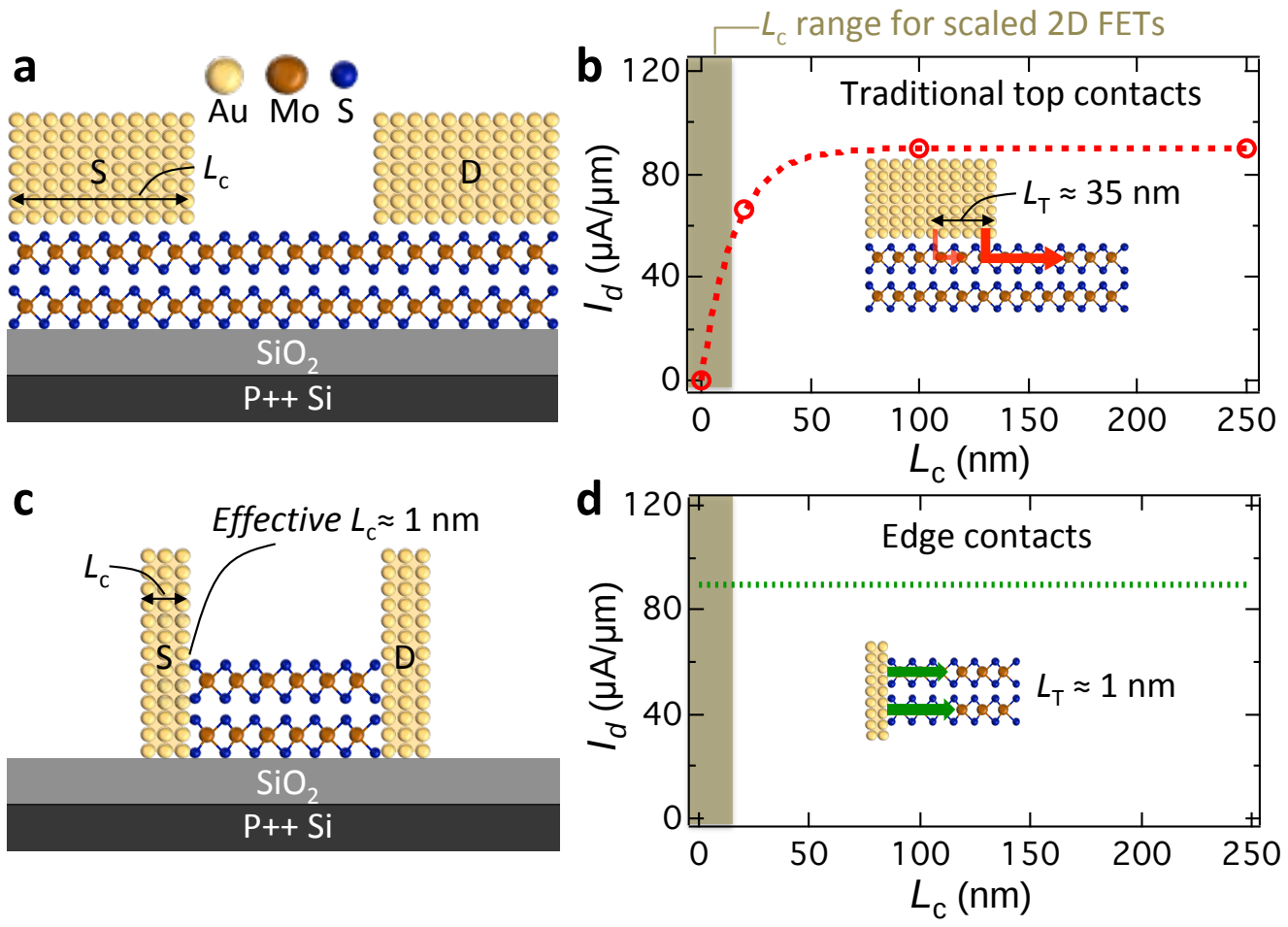

**Fig. 1 Top versus edge contacts to 2D $MoS_2$. a**, Schematic of a bilayer 2D FET with traditional top contacts. **b**, On-current diminishes as the top contact length decreases (data from ref.[28]), presenting a major roadblock for aggressively scaled transistors. Transfer length is indicated in inset schematic. **c**, Schematic of a bilayer 2D FET with edge contacts and an effective $L_c$ < 1 nm, leading to the possibility of **(d)** on-current that is independent of contact length.

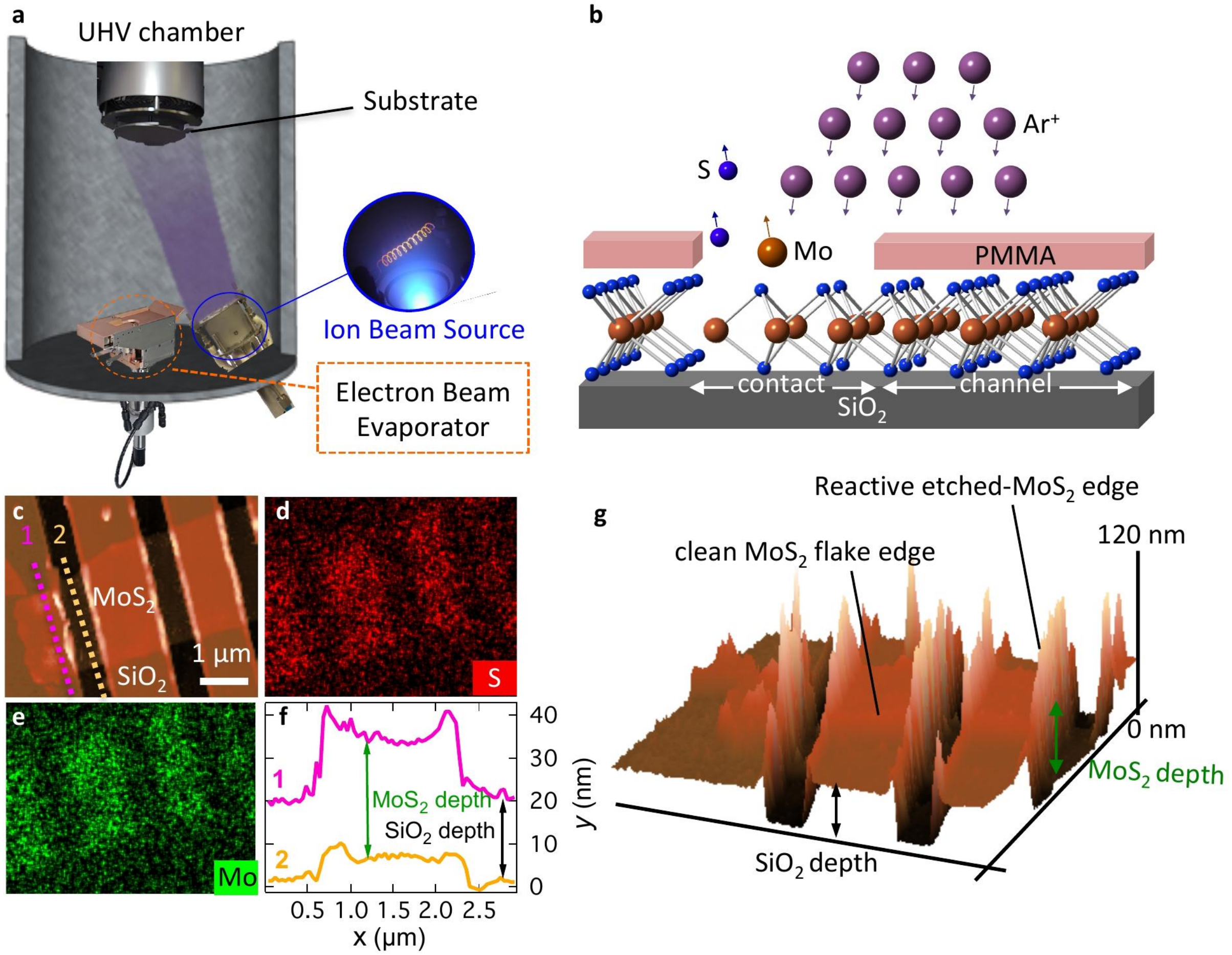

**Fig. 2. *In situ* etching of $MoS_2$.** **a**, Ion beam source and e-beam evaporator incorporated within the same UHV chamber. **b**, Schematic of the etch process with only contact regions selectively bombarded by Ar ion beam. **c**, AFM image of $MoS_2$ flake after etching and PMMA removal. EDS mapping of flake in **c** gives the sulfur signal in **d** and molybdenum signal in **e**. **f**, Line scan height profiles 1 and 2 from the AFM image in **c**. **g**, 3D AFM image of **c** highlighting the reactive etched $MoS_2$ edges and the relatively clean $MoS_2$ flake edges.

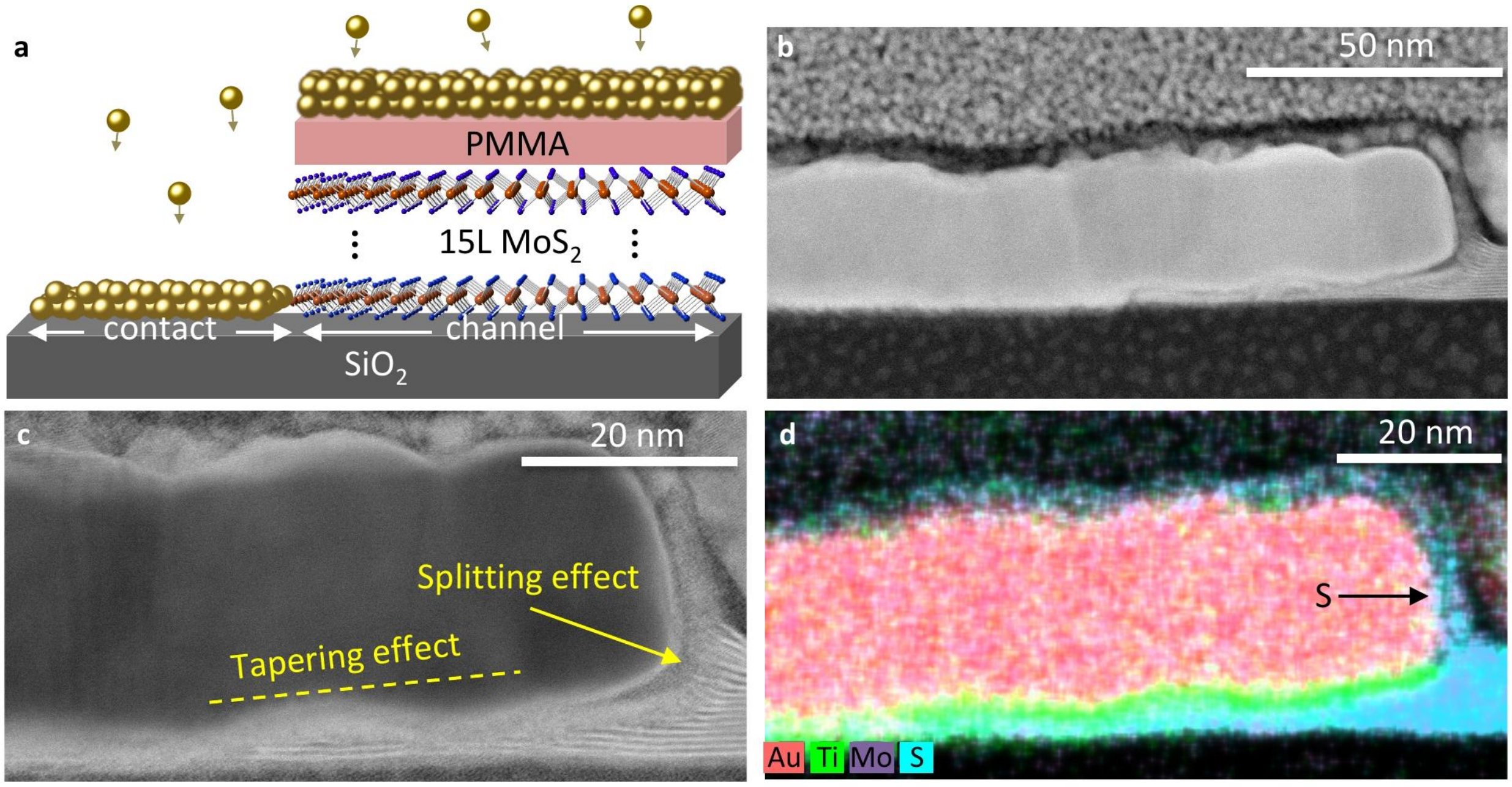

**Fig. 3. Metal-$MoS_2$ edge interface**. **a**, Diagram of the *in situ* metal deposition process forming an edge contact for 15L $MoS_2$ flake with 2 nm Ti / 20 nm Au. **b**, Cross-sectional STEM image of $L_c$ = 200 nm contact. **c**, Magnification of left edge of the contact showing tapering and splitting effects. **d**, EDS image of right side of the contact mapping the presence of different elements.

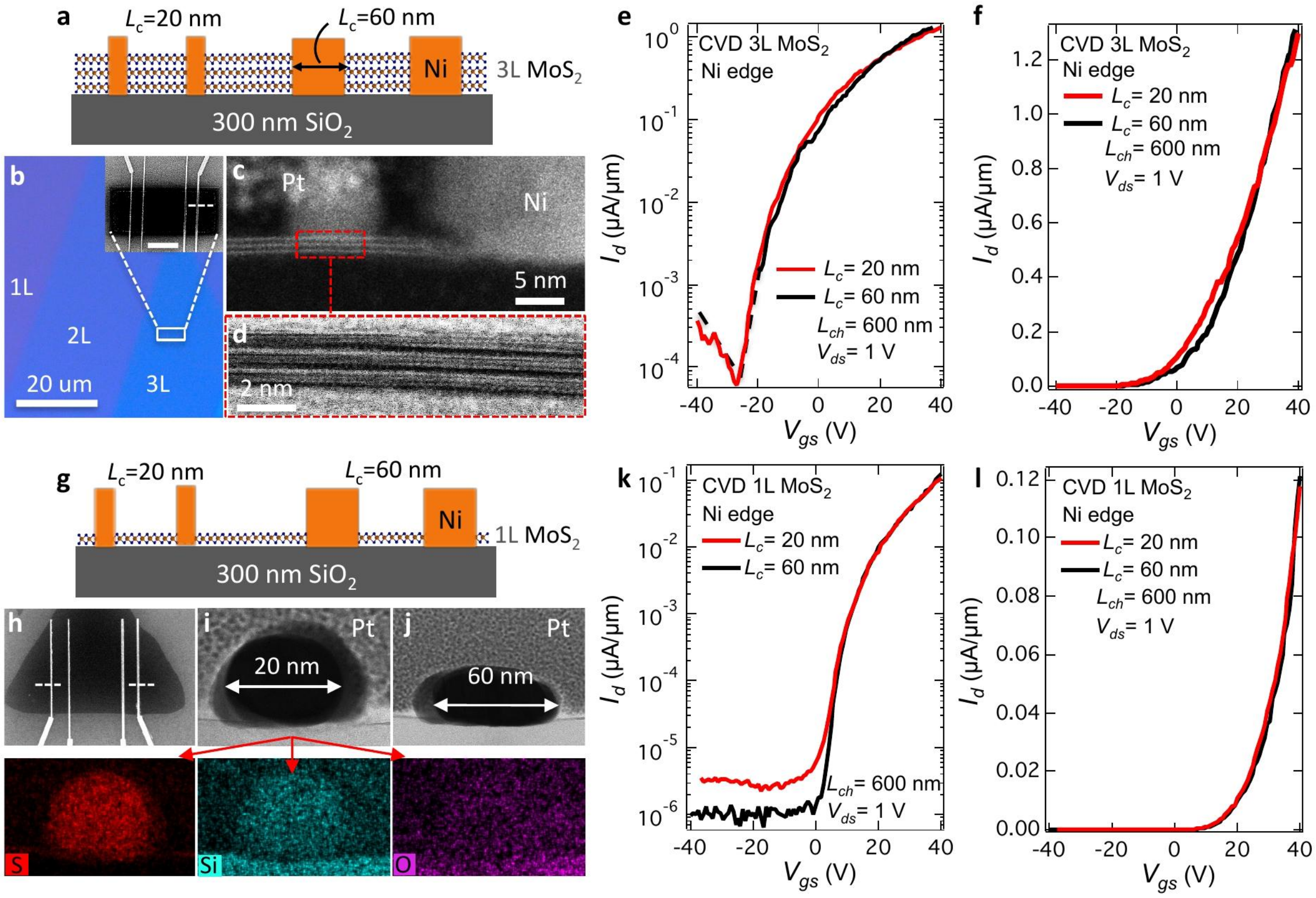

**Fig. 4. Trilayer and monolayer $MoS_2$ FETs with Ni edge contacts**. **a**, Schematic of edge-contacted devices on 3L $MoS_2$. **b**, Optical image of CVD-grown flakes with inset SEM image of trilayer $MoS_2$ FETs; scale bar in SEM image is 1 μm. Cross-sectional STEM images of: **c,** right edge of $L_c$=60 nm contact and **d**, atomic side-view of the trilayer $MoS_2$. **e,** Subthreshold and **f**, transfer characteristics of the edge-contacted devices, showing performance that is independent of contact length. **g**, Schematic of edge-contacted devices on monolayer $MoS_2$. **h**, SEM image of the devices with a scale bar of 1 μm. STEM images of: **i,** $L_c$ = 20 nm contact and **j,** $L_c$ = 60 nm contact. Arrows point to corresponding EDS scans of sulfur, silicon, and oxygen in **i**. **k,** Subthreshold and **l**, transfer characteristics of the monolayer edge-contacted devices, also showing the performance that is independent of contact length.

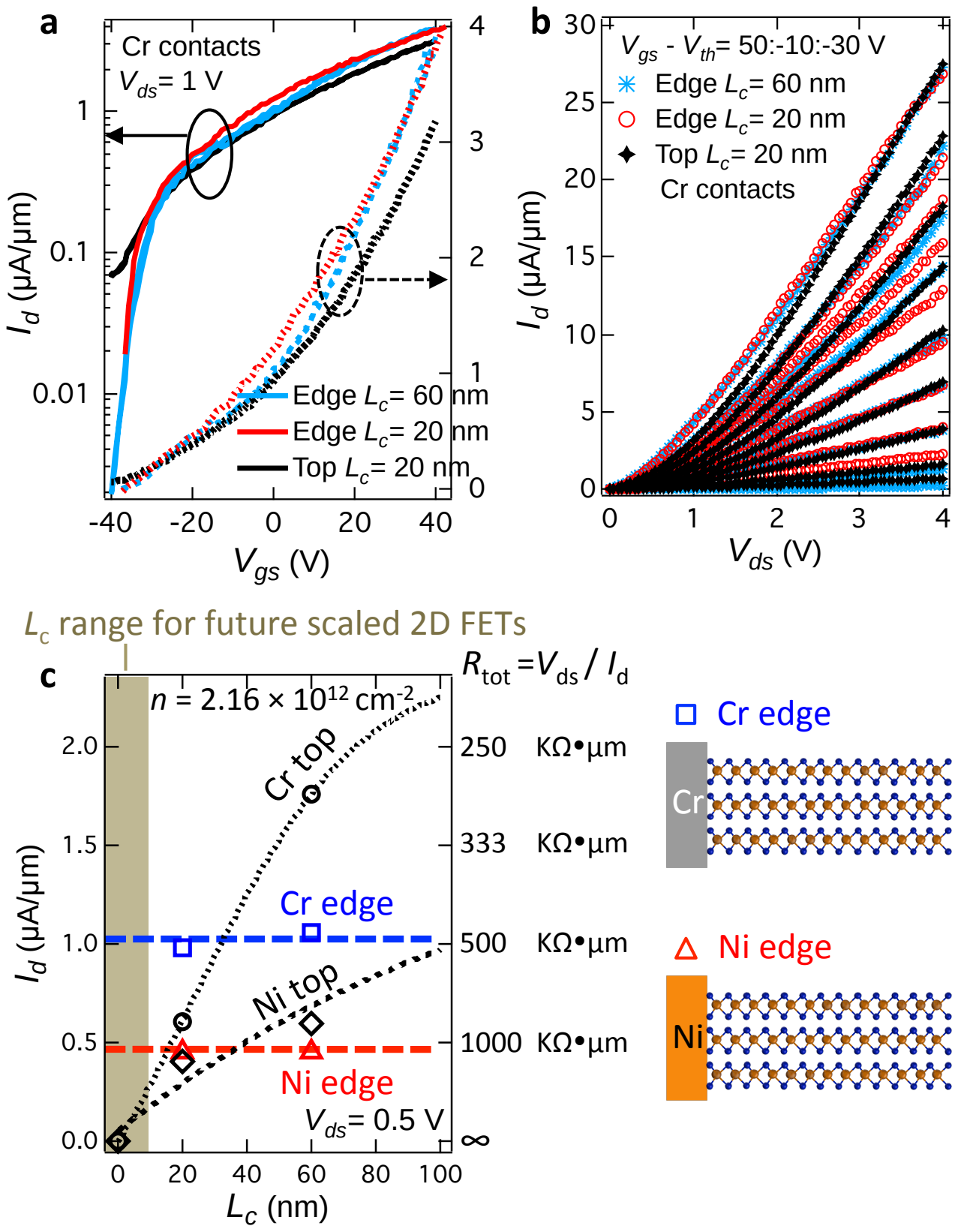

**Fig. 5. Ultimate scaling of contact length**. Comparison of Cr top and edge contacts to $MoS_2$ FETs with **a,** subthreshold & transfer and **b**, output curves. **c**, Relationship between $I_d$ and $L_c$ for different contact schemes, showing the advantage of edge contacts, especially in the short contact length region. The total resistance $R_{tot}$ was listed on the right side with $R_{tot}$ = 0.5 V/ $I_d$. The $L_{ch}$ for all the top- and edge-contacted devices are 600 nm, which suggests the $R_{ch}$ should be similar.

# Supplementary Information

## 1. Supplementary Notes

**Supplementary Note 1: Partial-edge contacts to exfoliated multilayer $MoS_2$**

The transistors in Supplementary Notes Figure 1 were built on an exfoliated $MoS_2$ flake with a thickness of ~35 nm, based on the AFM measurement included in the inset. These devices were etched for 150 s (etch-depth of ~28 nm), resulting in partial edge exposure of the $MoS_2$ in the contact area, as depicted in Supplementary Notes Figure 1b. The contact metal and the remaining layers in the center of the contact region could act as the additional carrier injection route, forming essentially a top contacts (see Supplementary Fig. 5(a)). Hence, carrier injection through both the top and edge are present, which is the reason why the contacts here are partial edge contacts. The *I-V* characterizations of the partial edge-contacted devices were given in Supplementary Notes Figure 1(c-e), showing an on-current level of 9 μA/μm at $V_{ds}$=1 V, which is on par with other top contacted devices in the literature[1].

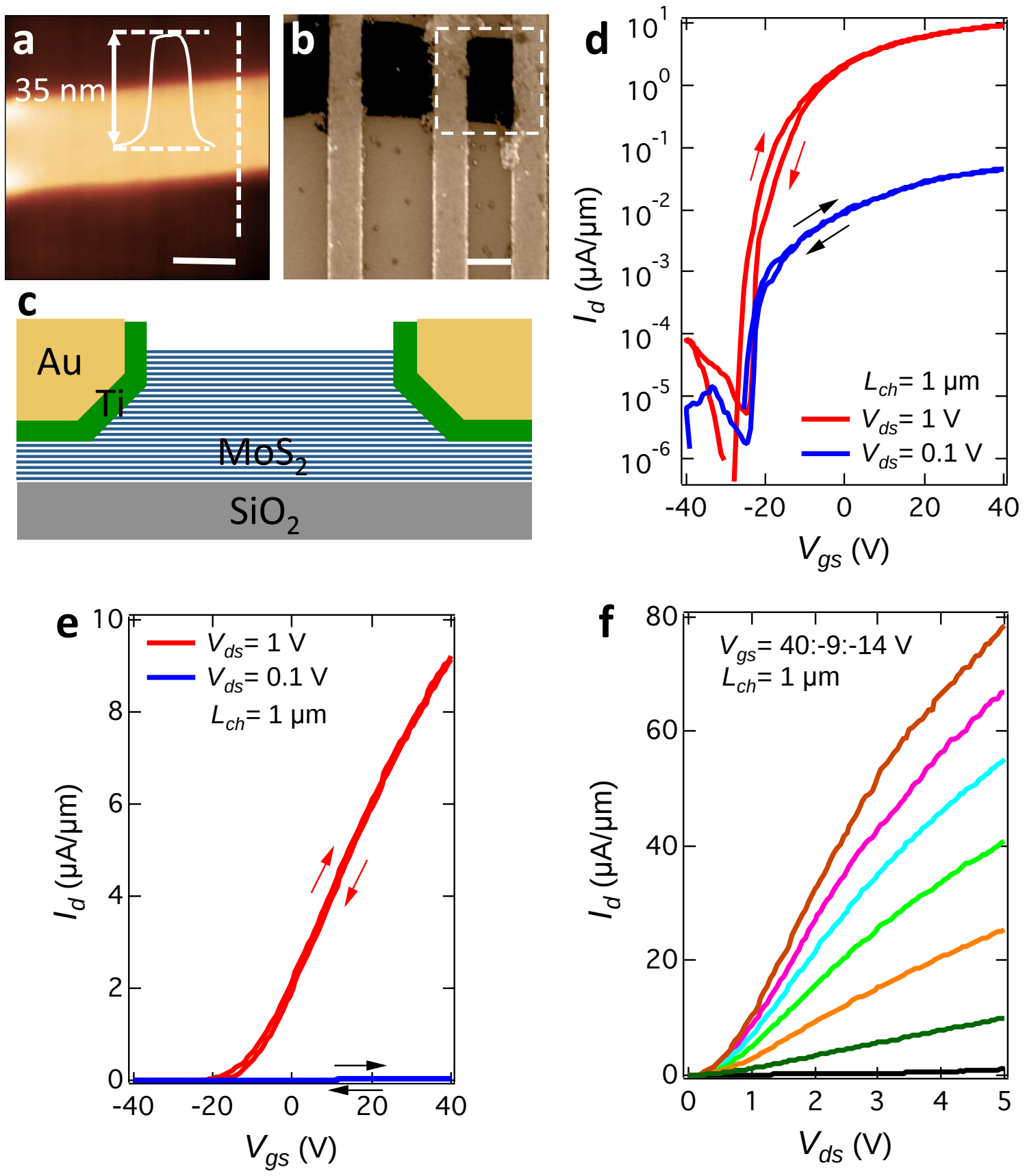

**Supplementary Notes Figure 1**. **Partial-edge contacts to exfoliated multilayer $MoS_2$. a**, AFM image of the exfoliated flake with a thickness of 35 nm. **b**, SEM image and **c**, schematic of a partial edge-contacted device in the dashed box in (**b**). The scaled bar in both (**a**) and (**b**) is 1 µm. $I_d$-$V_{gs}$ (**d**) subthreshold and (**e**) transfer curves of the device tested under ambient conditions. **f**, $I_d$-$V_{ds}$ curves of the device showing rectifying behavior when $V_{ds}$ is below 1 V.

**Supplementary Note 2: Quasi-edge contacts to exfoliated multilayer $MoS_2$**

Transistors in Supplementary Notes Figure 2a were built on the same flake as the devices in Supplementary Note 1. The flake was etched for 260 s (etch-depth of ~48 nm), producing quasi-edge contacts to $MoS_2$ with tapering and splitting effect but without apparent top contacting layers, as depicted in Supplementary Notes Figure 2b. Due to the thickness of the $MoS_2$ flake and substrate-gated device structure, the majority of the current is injected into the $MoS_2$ near the bottom of the flake, and thus injection at the contacts is likely to dominate near the bottom of the contact region. Performance of the quasi-edge contacted devices is shown in Supplementary Notes Figure 2(c-d). Two $I_d$-$V_{gs}$ sweeps are shown from the same $V_{ds}$= 1 V, where the 1$^{st}$ sweep is the initial measurement and the 2$^{nd}$ is taken after a sweep at $V_{ds}$= 3 V was performed. The 6.5x jump from 0.79 $\mu$A/$\mu$m to 5.17 $\mu$A/$\mu$m between these two sweeps suggests a forming or "burn-in" effect at the quasi-edge contacts, creating more favorable bonds between the $MoS_2$ edge states and the metal. Another interesting observation is the threshold voltage shift from $V_{th}$ = -15 V in the partial to $V_{th}$ = -35 V in the quasi-edge contact case. It is also observable that the current remains approximately constant for $V_{gs}$ = 40 V to -20 V in the first sweep (blue curve), which is not the case for the devices in the partial-edge contacts scheme. These differences suggest a distinction in the carrier injection behavior and gating effect in the partial- versus quasi-edge contacts to exfoliated multilayer $MoS_2$.

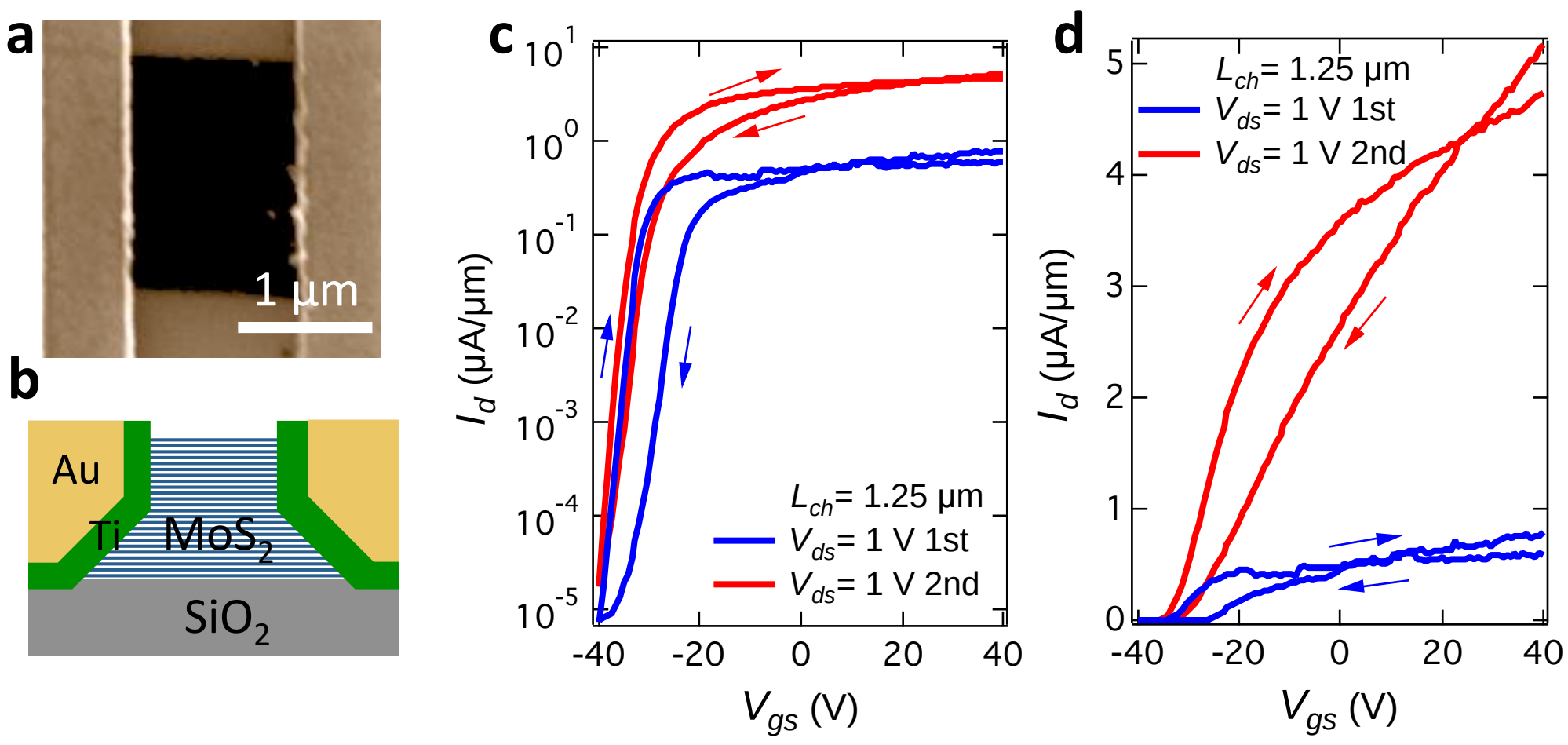

**Supplementary Notes Figure 2**. **Quasi-edge contacts to exfoliated multilayer $MoS_2$ (35 nm thick). a,** SEM image and **b**, schematic of a quasi-edge contacted device. $I_d$-$V_{gs}$ (**c**) subthreshold and (**d**) transfer curves of the device tested under ambient.

Quasi-edge contacts to a thinner flake were also demonstrated. Transistors in Supplementary Notes Figure 3a were built on an 8 nm thick $MoS_2$, as shown in the AFM image in Supplementary Notes Figure 3b. The flake was etched for 150 s (etch-depth of ~28 nm), producing quasi-edge contacts to $MoS_2$, as depicted in Supplementary Notes Figure 3c. Considering the thickness of this flake is similar to the 10 nm flake demonstrated in the manuscript, we expect the tapering effect to show up here. Performance of the quasi-edge contacted devices is shown in Supplementary Notes Figure 3(d-e). Two $I_d$-$V_{gs}$ sweeps are shown from the same $V_{ds}$= 1 V, where the 1$^{st}$ sweep is the initial measurement and the 2$^{nd}$ is taken after a sweep at $V_{ds}$= 5 V was performed. The 60x jump from 0.02 $\mu$A/$\mu$m to 1.2 $\mu$A/$\mu$m between these two sweeps suggests a "burn-in" or forming effect similar to the one in Supplementary Notes Figure 2. Compared to the quasi-edge contacted devices in Supplementary Notes Figure 2, the quasi-edge contacted devices with 8 nm flake thickness have a smaller current, which may attribute to the thinner flake thickness, which leads to the smaller edge contact area.

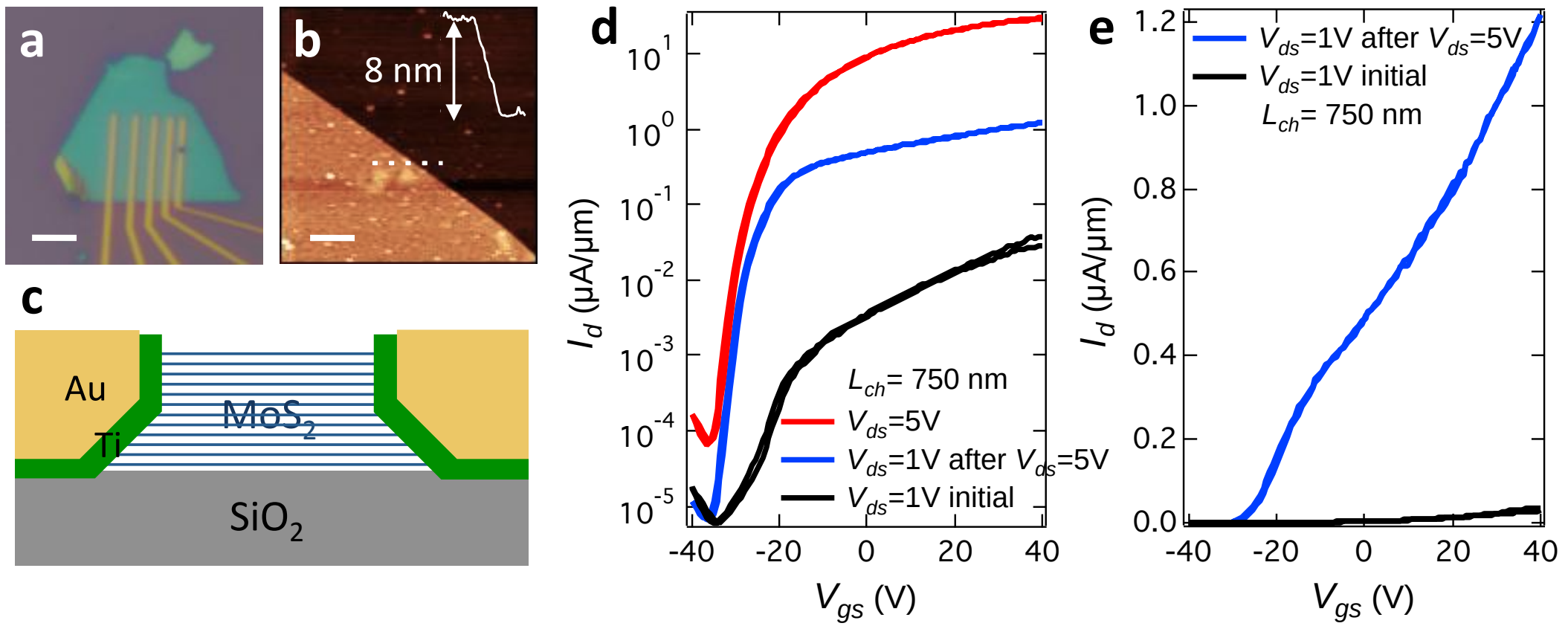

**Supplementary Notes Figure 3**. **Quasi-edge contacts to exfoliated multilayer $MoS_2$ (8 nm thick). a,** SEM image (scale bar, 2 μm) and **b**, AFM image showing 8 nm of flake thickness (scale bar, 500 nm). **c**, schematic of a quasi-edge contacted device. $I_d$-$V_{gs}$ (**d**) subthreshold and (**e**) transfer curves of the device tested under ambient.

**Supplementary Note 3: Benchmarking performance for top and edge contacts**

It can be seen from Supplementary Table 1 that most of the CVD-based $MoS_2$ devices have a very large contact resistance (30~200 kΩ*μm) and small $I_{on}$ compared to the exfoliated flakes[2–4]. Several papers using the same metal top contacts report drastically different contact resistances, which indicates the quality of the CVD-grown films, the deposition condition of the metal contacts and the fabrication process can all play a role in determining the final device performance. Also note that the different papers cited in Supplementary Table 1 have different overdrive voltage ($V_{ov}$), carrier density ($n$), channel lengths ($L_{ch}$), and number of layers ($N_L$). These factors need to be considered in order to have a fair comparison.

A few factors can explain why the $R_{tot}$ of the top and *in situ* edge contacts in this work is larger than the first few top contacts with higher performance listed in Supplementary Table 1. First, small length of $L_c$ is used with only $L_c$ = 20 nm (effective $L_c \approx$ 1 nm) for the edge contacts and $L_c \leq$ 60 nm for top contacts. Second, we use relatively low $V_{ov}$= 30 V on 300 nm thick $SiO_2$, which means lower carrier density ($n$ = $2.16\times10^{12}$ $cm^{-2}$) compared to the ones used by others in the Supplementary Table 1. The lower carrier density increases the $R_{sh}$ for all devices and $R_c$ for top contacts. Finally, relatively poor quality of $MoS_2$ films grown on $SiO_2$ (see Supplementary Fig. 9).

**Supplementary Table 1:** Benchmarking contacts for transistors using CVD-grown $MoS_2$

| **Ref.** | **Contact Strategy** | **EOT** nm | $V_{ov}$ V | $n$ $10^{12}cm^{-2}$ | $I_{on}$ µA/um | $I_{on}$ @ $V_{ds}$ (V) | $I_{on}/I_{off}$ $10^x$ | $L_{ch}$ µm | $R_c$ KΩ•µm | $N_L$ | $L_c$ µm |
|---|---|---|---|---|---|---|---|---|---|---|---|
| 5 | Ag/Au | 30 | 25 | 17.98 | 18.75 | 1 | 7 | 4.3 | **~3** | 1L | 4 |
| 6 | Cr/Pd | 300 | 100 | 7.19 | 8.97 | 0.5 | 8 | 1 | $R_{tot}$=55.7 | 2L | ~1 |
| 7 | UHV Au | 90 | 21 | 5.03 | 35 | 1 | 4 | 0.2 | **6.5** | 1L | 0.67 |
| 8 | Ti/Au | 285 | 125 | 9.47 | 9.00 | 1 | 6 | 1 | **20** | 1L | - |
| 9 | Graphene | 5~9 | - | - | 0.14 | 0.025 | - | 22 | **~30** | 1L | 35 |
| 10[*§] | Graphene overlap | 300 | 80 | 5.75 | 2.40 | 4 | 6 | 12 | **100** | 1L | 10 |
| 8[§] | Ti/Au | 6.4 | 7 | 23.72 | 1.75 | 1 | 3 | 1 | **175** | 1L | - |
| 11 | Au | 285 | 60 | 4.55 | 0.38 | 0.1 | 4 | 1 | **210** | 1L | 1 |
| 12 | Graphene overlap | 300 | 40 | 2.88 | 1.50 | 1 | - | 8 | **300** | 1L | 1.14 |
| 13 | Ni/Ti/Au | 300 | 80 | 5.75 | 0.27 | 1 | 5 | 10 | $R_{tot}$=3740 | 2L | 35 |
| 14 | Sc/Ni edge to hBN/$MoS_2$/hBN (*ex situ*) | 285 | 55.5 | 4.16 | 0.114 | 0.5 | 4 | 1.8 | $R_{tot}$=4386 | 1L | ~1 |
| 15[¶] | Ti/Au edge to hBN capped $MoS_2$ (*ex situ*) | 285 | 80 | 6.06 | 0.047 | 1 | - | N/A | $R_{tot}$=21M Ω | 1L | - |
| **This work** | Cr edge (*in situ*) | 300 | 30 | 2.16 | 1 | 0.5 | 4 | 0.6 | $R_{tot}$=500 $R_c$=220 | 3L | 0.02 |
| **This work** | Cr edge (*in situ*) | 300 | 30 | 2.16 | 0.8 | 0.48 | 4 | 2.2 | $R_{tot}$=600 $R_c$=190 | 3L | 0.1 |

$V_{ov} = V_{gs}-V_{th}$.

The data for the first *in situ* Cr edge contacts is from Fig. 5 of the main manuscript, whereas data for the second *in situ* edge contacts comes from Supplementary Notes Figure 4.

[*] It is unclear how large the overdrive voltage is. But the carrier density is high, $n$ ~ $1\times10^{13}$ $cm^{-2}$. A back-gated device is also reported within the paper with 300 nm $SiO_2$ as gate dielectric, 70 V as overdrive voltage ($5.03\times10^{13}$ $cm^{-2}$), but the resulting $I_d$ is only 0.09 µA/um at $V_{ds}$=1 V. The estimated $R_{tot}$ = 11.11 MΩ•µm.

[§] These reports use top gate structure and all other reports use back gate structure.

[¶] The author also demonstrated Pd/Au edge contacts to hBN-encapsulated $MoS_2$ but with $R_{tot}$ = 333 MΩ. Other metals edge contacts such as Ti/Au (0.5/50 nm) and Al/Cr/Au (40/10/30 nm) is close to open circuit. Note that $I_d$ and $R_{tot}$ in this reference is not normalized to contact width. In this row, $R_{tot}$ was calculated by using 1 V/$I_d$. Because of the S-shape output characteristics, the actual $R_{tot}$ is likely even larger.

From Supplementary Figure 10, we can extract the $R_{sh}$ = 100 KΩ/square for 3L $MoS_2$ at the carrier density of $n = 2.16 \times 10^{12}$ $cm^{-2}$ ($V_{ov}$ = 30 V on 300 nm $SiO_2$). Using this $R_{sh}$, the data from Supplementary Note Fig. 4, and the Supplementary Fig. 6 and 12, we estimated the contact resistance for various devices in the following Supplementary Table 2. Au edge contacts have a very large contact resistance compared to other metals, indicating a poor Au-$MoS_2$ edge bonding. Both Ni and Cr edge contacts outperform their top-contacted counterparts at scaled dimension ($L_c$ = 20 nm).

**Supplementary Table 2:** Top vs. Edge contacts to CVD-grown 3L $MoS_2$

| **Metal** | **$R_c$ for top contacts ($L_c$=60 nm)** | **$R_c$ for top contacts ($L_c$=20 nm)** | **$R_c$ for edge contacts** |
|---|---|---|---|
| Au | 164 KΩ•μm | N/A | > 10 MΩ•μm |
| Ni | 386 KΩ•μm | 595 KΩ•μm | 525 KΩ•μm |
| Cr | 110 KΩ•μm | 381 KΩ•μm | 205 KΩ•μm* |

The extraction of $R_c$ is at $V_{ds}$ = 0.5 V and overdrive voltage ($V_{ov}$) of 20 V for Au, 30 V for Ni and Cr. The $MoS_2$ film thickness used in these devices is 3 layers. These $MoS_2$ film are grown on $SiO_2$ without any transfer process.

* The mean value of the last two rows from Supplementary Table 1, (190 KΩ•μm + 220 KΩ•μm)/2 = 205 KΩ•μm.

**Supplementary Note 4: Comparison of $L_T$, $R_c$ components for top and edge contacts**

$L_T$ is different in top and edge contacts. For a top contacts, according to the transfer length method, $L_T \approx (\rho_c/R_{sh})^{1/2}$, where $\rho_c$ is the contact resistivity of the top metal-$MoS_2$ interface and $R_{sh}$ is the sheet resistance of the $MoS_2$ underneath the top meal, as shown in the diagram in Supplementary Table 3. Theoretically, we can extract the $L_T$ from the x-intercept of a transfer length method (TLM) plot. For example, the x-intercept in Supplementary Fig. 10(d) suggests that $L_T$ for top Cr contacts is ~ 2.37 μm, which is significantly higher than the $L_T \approx$ 35 nm demonstrated in ref. 4. This estimation of $L_T$ is reasonable considering the difference of $R_c$ between this work and ref. 4. According to the aforementioned $L_T$ equation, we also expect $L_T$ for Ni top contacts to be larger than Cr top contacts, which is based on the fact that the $\rho_c$ of Ni-$MoS_2$ interface is larger than its Cr-$MoS_2$ counterparts (see Supplementary Table 2), with $R_{sh}$ to be the same since they all have the same $V_{ov}$. These analyses make us postulate the $I_d$ versus $L_c$ in Fig. 5(c) of the main manuscript should be linear in the range of $L_c$ < 100 nm. For edge contacts, because the carriers are injected completely through the edge, the $L_T$ length is the length of edge interface, which is about 1 nm, which sets the edge contacts apart from top contacts and makes the edge contact immune to contact scaling.

As mentioned in the manuscript, the contact resistance for edge contacts is intrinsically different compared to top contacts. In a traditional top contacts, the contact resistance includes the top-interfacial resistance of metal-$MoS_2$, plus the series resistance of the $MoS_2$ beneath the metal contacts (lateral carrier flow beneath the metal contacts), as shown in Supplementary Table 2. In the case of a multilayer film, the interlayer resistance also affects the contact resistance; but for the sake of simplicity, we excluded the interlayer resistance for top contacts in Supplementary Table 2. For edge contacts, however, the contact resistance is exclusively the resistance between the metal and the edge states of the 2D materials. Considering that the edge states will alter the bandgap at the termination of the 2D materials, the exact band diagram remains to be investigated using Density Function Theory.

**Supplementary Table 3:** Comparison of $L_T$, $R_c$ components, and band diagram between top and edge contacts

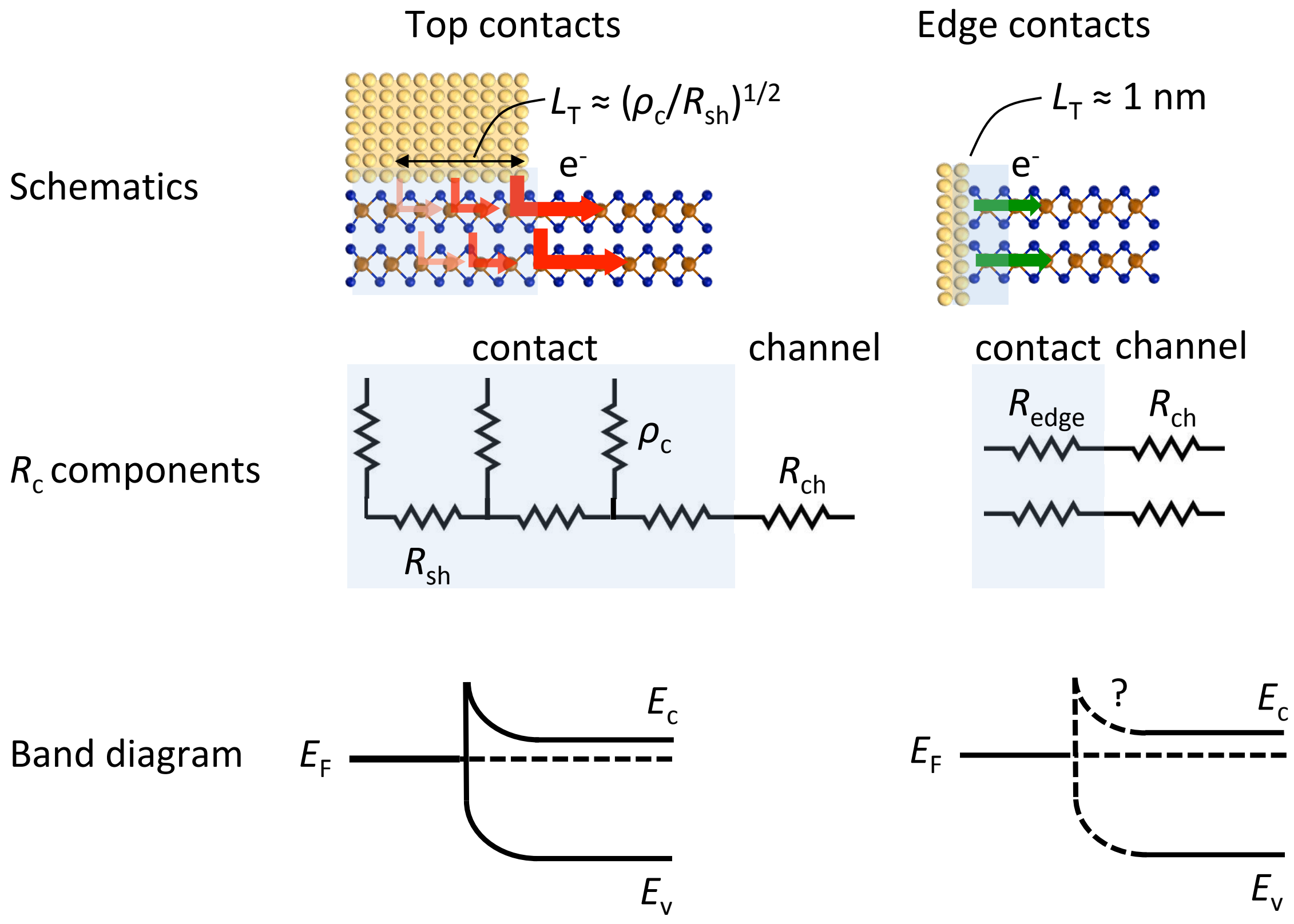

$R_{ch}$ can be approximated as $R_{sh}L_{ch}/W$, where $W$ is the width of contact electrodes.

**Supplementary Note 5: Low-temperature characterization of *in situ* Cr-$MoS_2$ edge contacts**

CVD-grown 3L-$MoS_2$ was used here. The etching condition was consistent with the one used in etching 3L $MoS_2$ for Ni edge contacts introduced in the Method section of the main manuscript. The device was characterized at temperatures ranging from 30 to 300 K. In the case of an electron-dominated transport, at low $V_{gs}$, high barrier shows up in the contact; there are limited carriers having the thermionic energy high enough to jump over the barrier to the drain side, yielding a low current flow. As the $V_{gs}$ increases, the barrier lowers. After moving over flat band condition, the barrier becomes thinner as $V_{gs}$ further increases, while the height of the barrier settles at $\phi_B$. The thinning of Schottky barrier introduces the tunneling component by carriers tunneling though the thin barrier. We also show the process using simple diagrams in Supplementary Note Fig. 4d.

According to the equation $I_d = AT^2 \exp((q\phi_B)/(K_B T))[1 - \exp((qV_{ds})/(K_B T))]$, the Schottky barrier height can be extracted. In this equation, $I_d$ is the current, $A$ is the Richardson's constant, $K_B$ is the Boltzmann constant, $q$ is the electronic charge, $T$ is the temperature, and $V_{ds}$ is the source to drain bias. After some mathematical transitions, the equation becomes $\ln(I_d/T^2) = \phi_B \bullet [q/(K_B \bullet T)]$. Plotting $\ln(I_d/T^2)$ on the y-axis and $(q/K_B)\bullet T$ on the x-axis makes a Arrhenius plot with the slope being $\phi_B$. For simplicity, some reports would put $1000/T$ on the x-axis, as demonstrated in Supplementary Note Fig. 5. A general guideline to interpret the Arrhenius plot is looking at the slop of the fitting curves for different $V_{gs}$ at different temperatures. When the fitting curves turn downward, which is indicative of a thermionic carrier transport over the Schottky barrier. If the fitting curves go up, then it suggests a tunneling transport through the Schottky barrier.

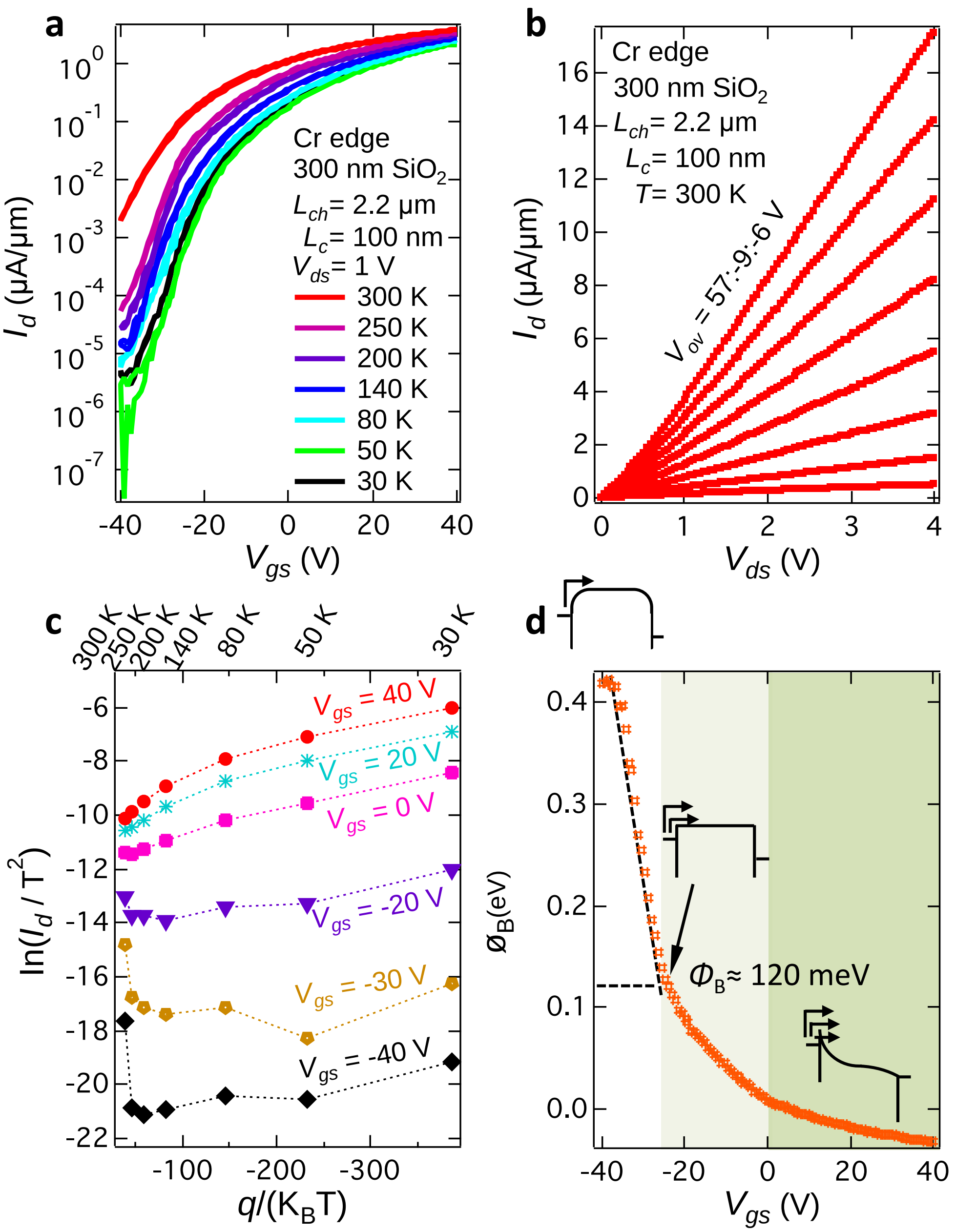

**Supplementary Notes Figure 4. Low-temperature characterization of *in situ* edge contacted $MoS_2$ transistor**. **a**, *I*-*V* characteristics of the multilayer Cr edge contacted devices across different temperature. **b**, Output curves of the device in room temperature. **c**, Arrhenius plot of the device. At low $V_{gs}$, the dashed fitting curves first go downward at high temperature (300-250 K), but the curves transition to flat and then go upward as the temperature drops below 200 K. **d**, Extracting Schottky barrier height of the device.

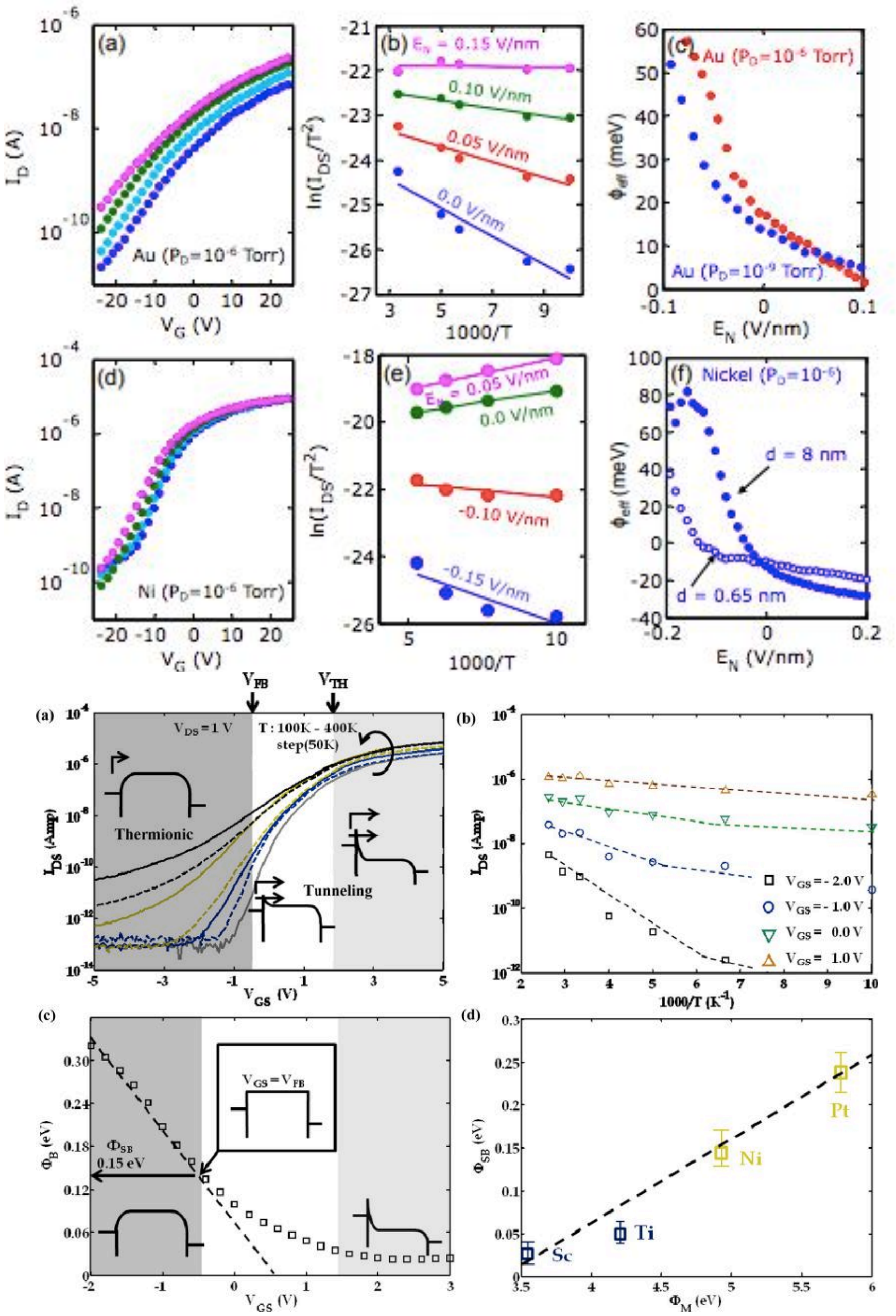

**Supplementary Notes Figure 5. Examples of low-temperature characterization of top contacted $MoS_2$ transistor**. The top six panels are adapted from ref. 4, with temperatures ranging from 100 to 200 K. The bottom four panels are adapted from ref. 17, with temperatures ranging from 150 to 350 K. It can be seen that the fitting curves in the Arrhenius plot all turn downward at low $V_{gs}$ while only go flat or up when the $V_{gs}$ because large enough so that the Schottky barrier becomes too thin and tunneling becomes dominant, which is in sharp contrast with the Arrhenius plot in Supplementary Note Fig. 4(c).

## Supplementary Figures

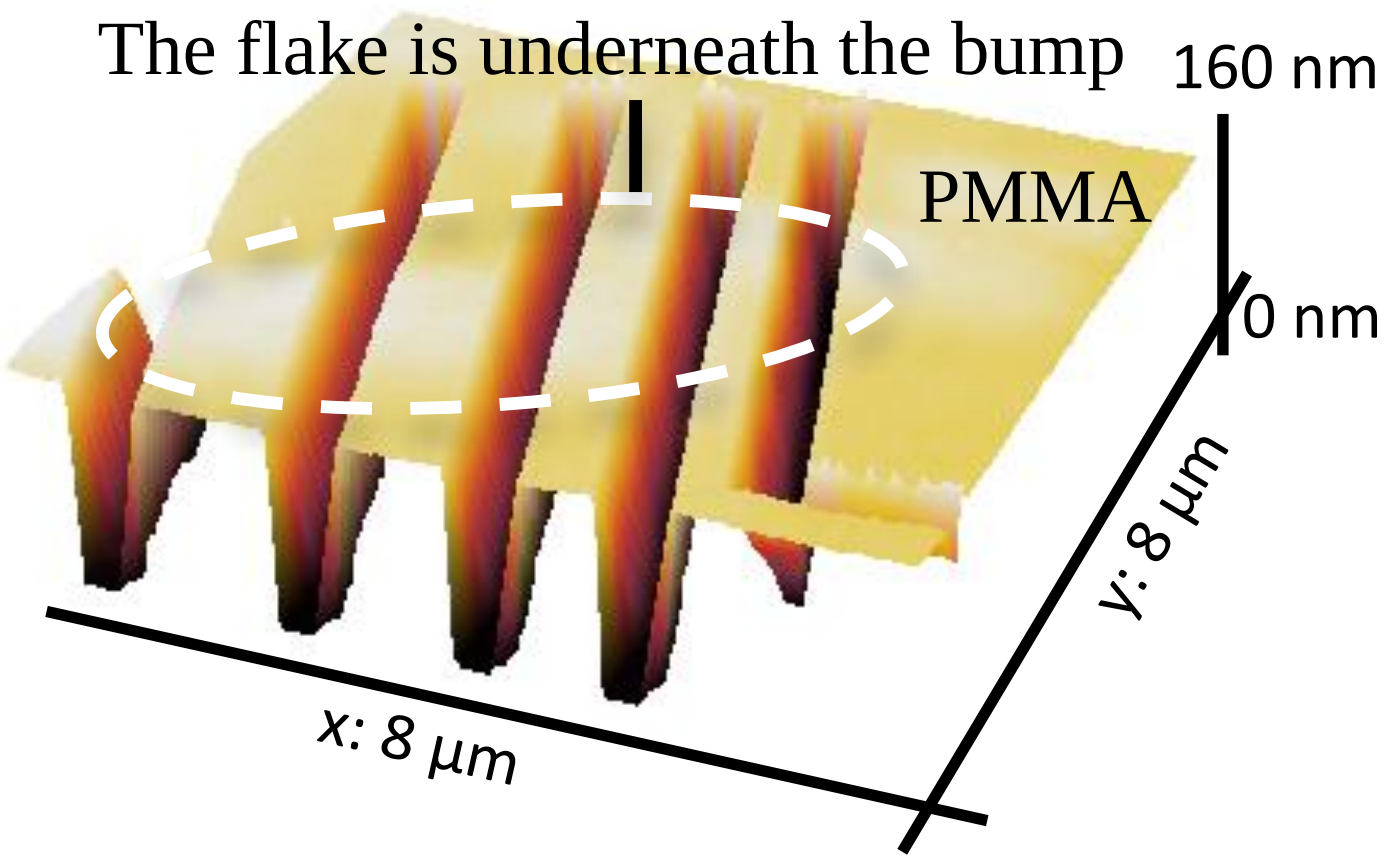

**Supplementary Figure 1. 3D AFM image of the flake in Fig. 2 of the main text covered with PMMA after ion beam etching**. Considering the smooth surface and clean edge of the PMMA, the etching effect of ion beam on PMMA can be neglected.

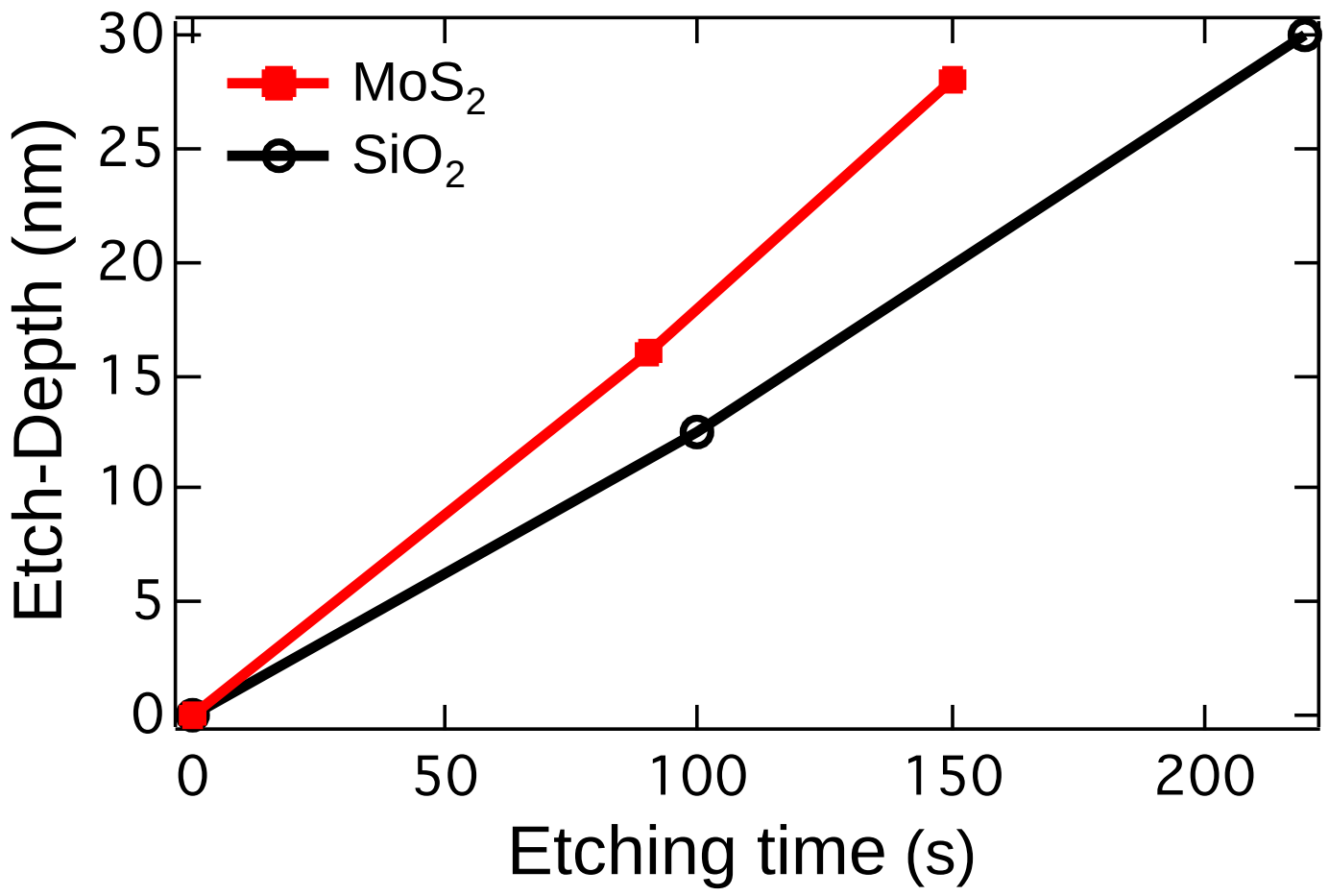

**Supplementary Figure 2. Etch-depth vs. etching time for $MoS_2$ and $SiO_2$**. The etching condition is 600 eV, 36 mA, and Ar ion beam.

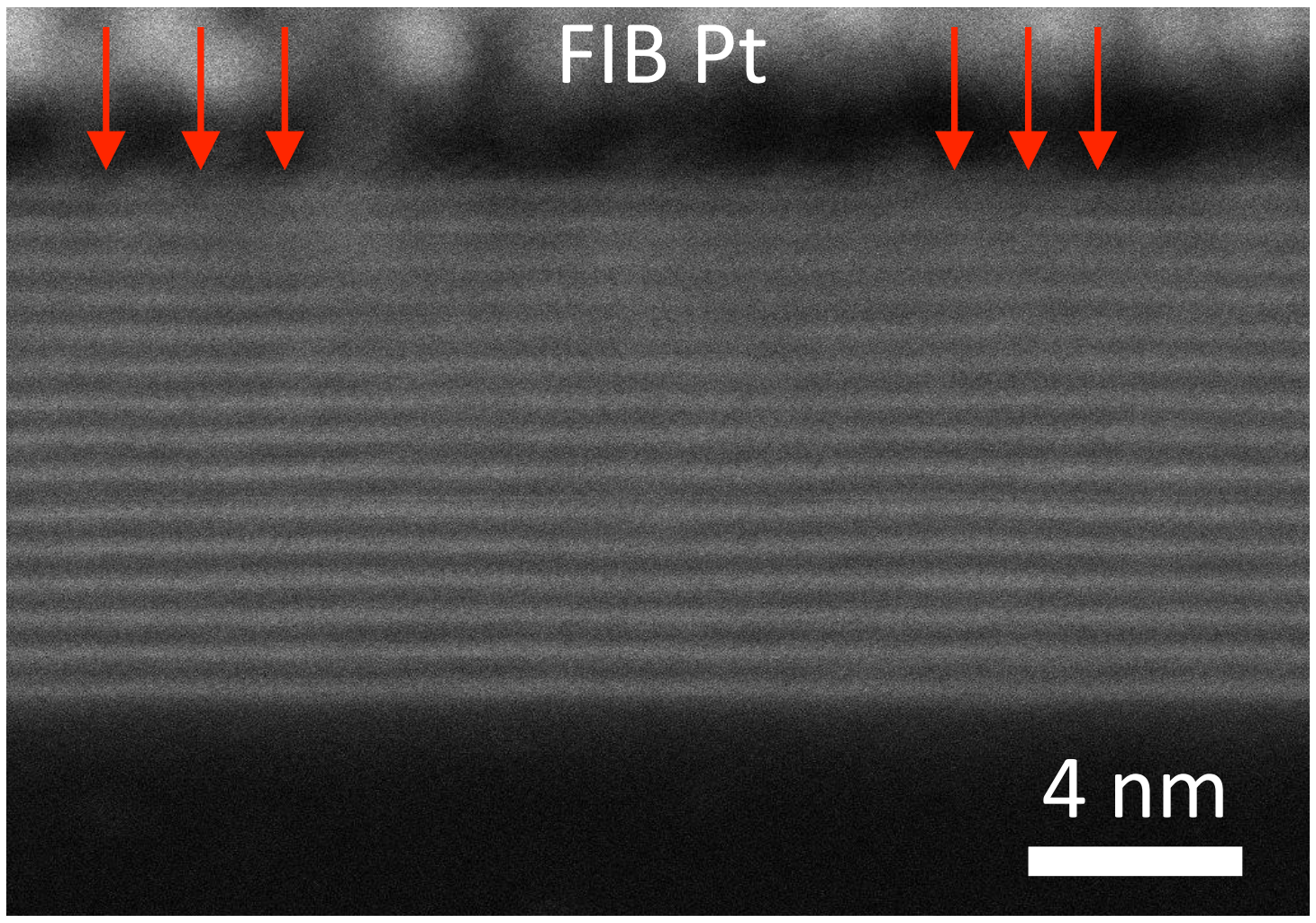

**Supplementary Figure 3. Cross-sectional image of the exfoliated $MoS_2$ flake used in Fig. 3**. The thickness of this flake is about 10 nm (15 layers). The metal on top of the $MoS_2$ layers is Pt, which protects the flake from FIB process. More defects seem to show up in the top layer identified by the arrows, whereas the layers underneath the top layer are more uniform.

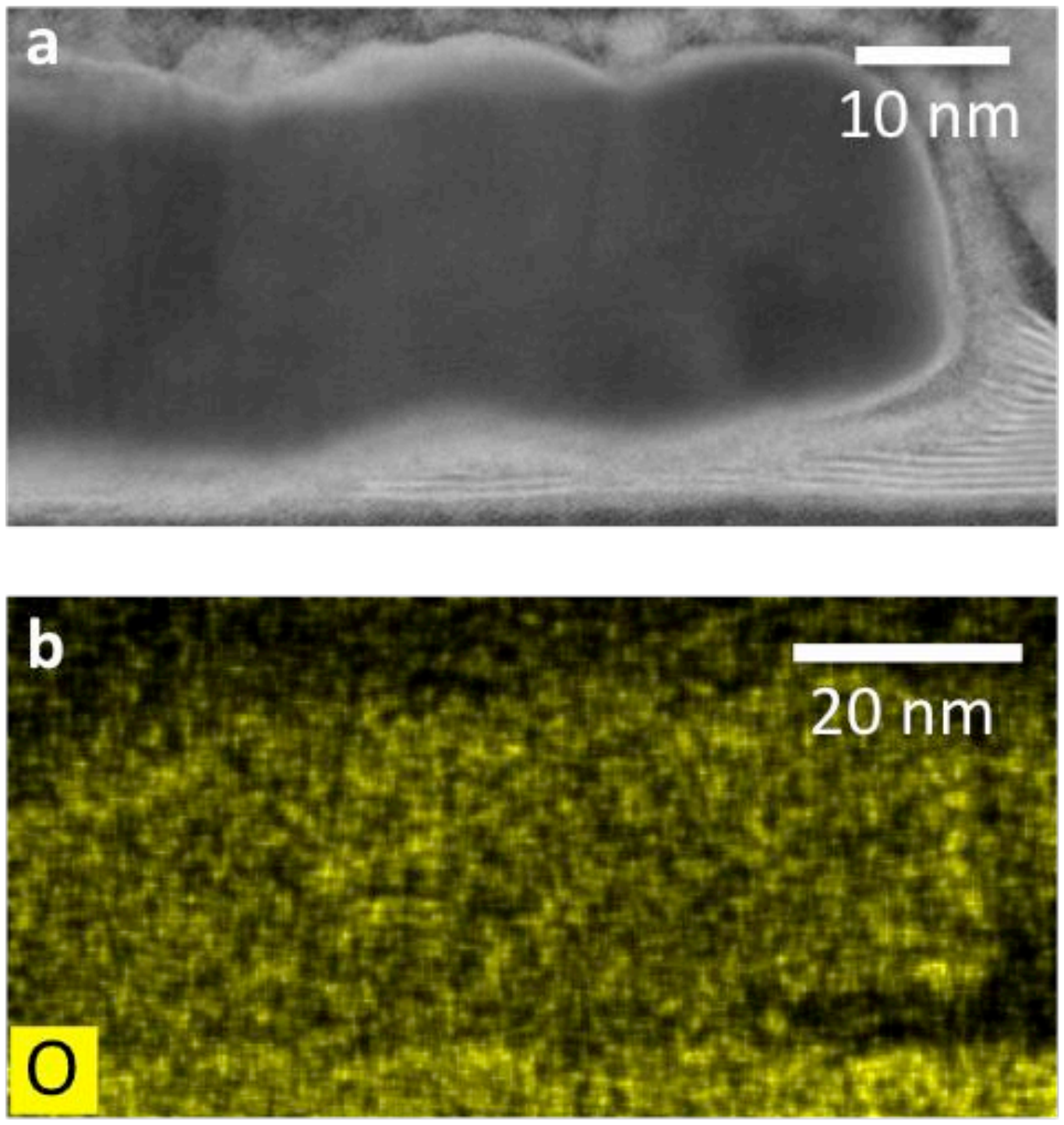

**Supplementary Figure 4. EDS mapping oxygen at the etched $MoS_2$-metal interface. a**, A magnified view of the etched $MoS_2$-metal interface. **b**, Mapping the oxygen signal at the interface. No excess of O element shows up at the $MoS_2$ edge-metal interface.

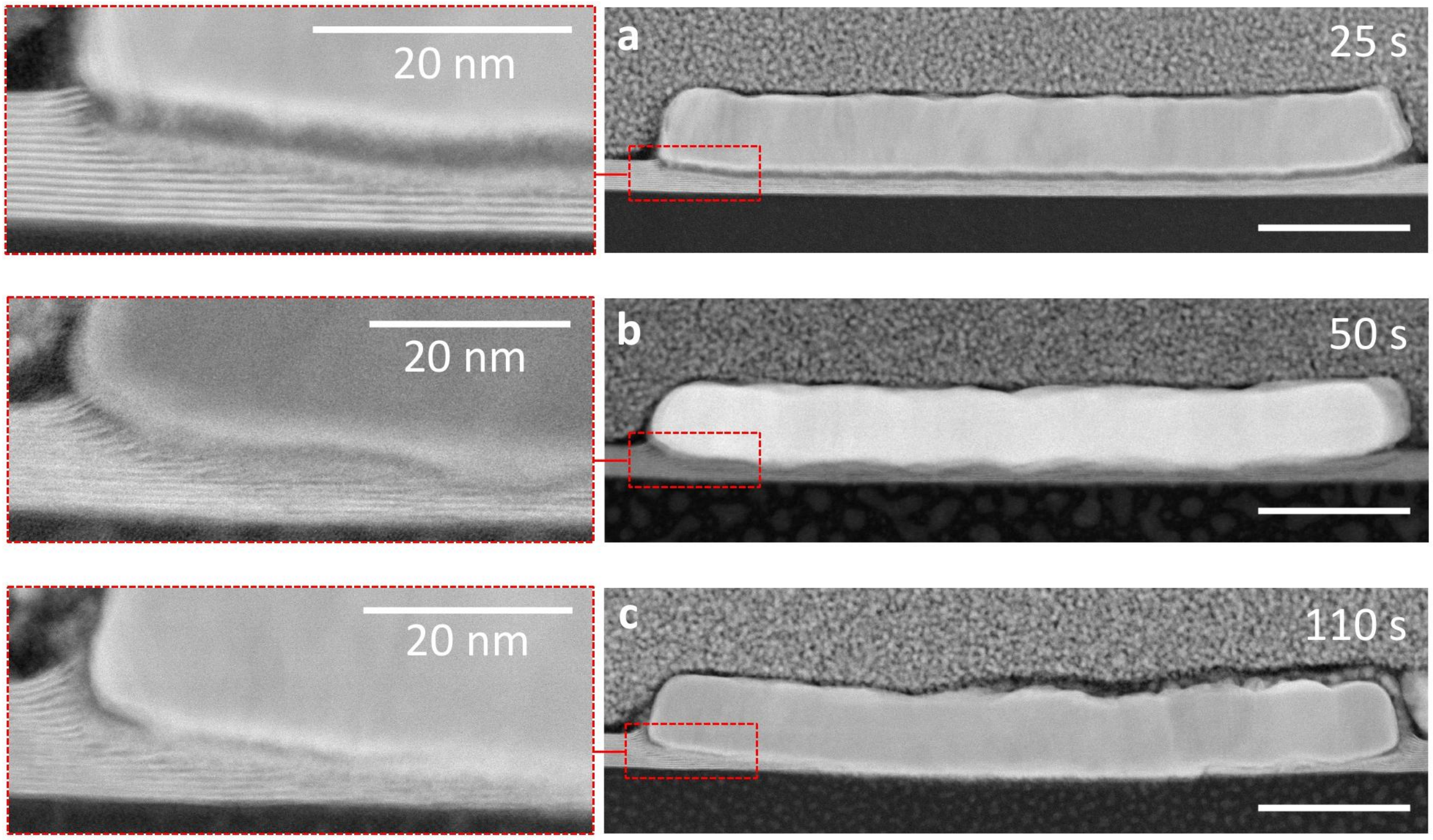

**Supplementary Figure 5**. **Effect of etching time**. Etching time of (**a**), (**b**), and (**c**) is 25 s, 50 s, and 110 s, respectively. Due to the tapering effect, the center region of the contact was etched faster than the edge region. A similar splitting effect can be seen on the zoom-in view of the left edges. The scale bar on the bottom right of (**a-c**) is 50 nm.

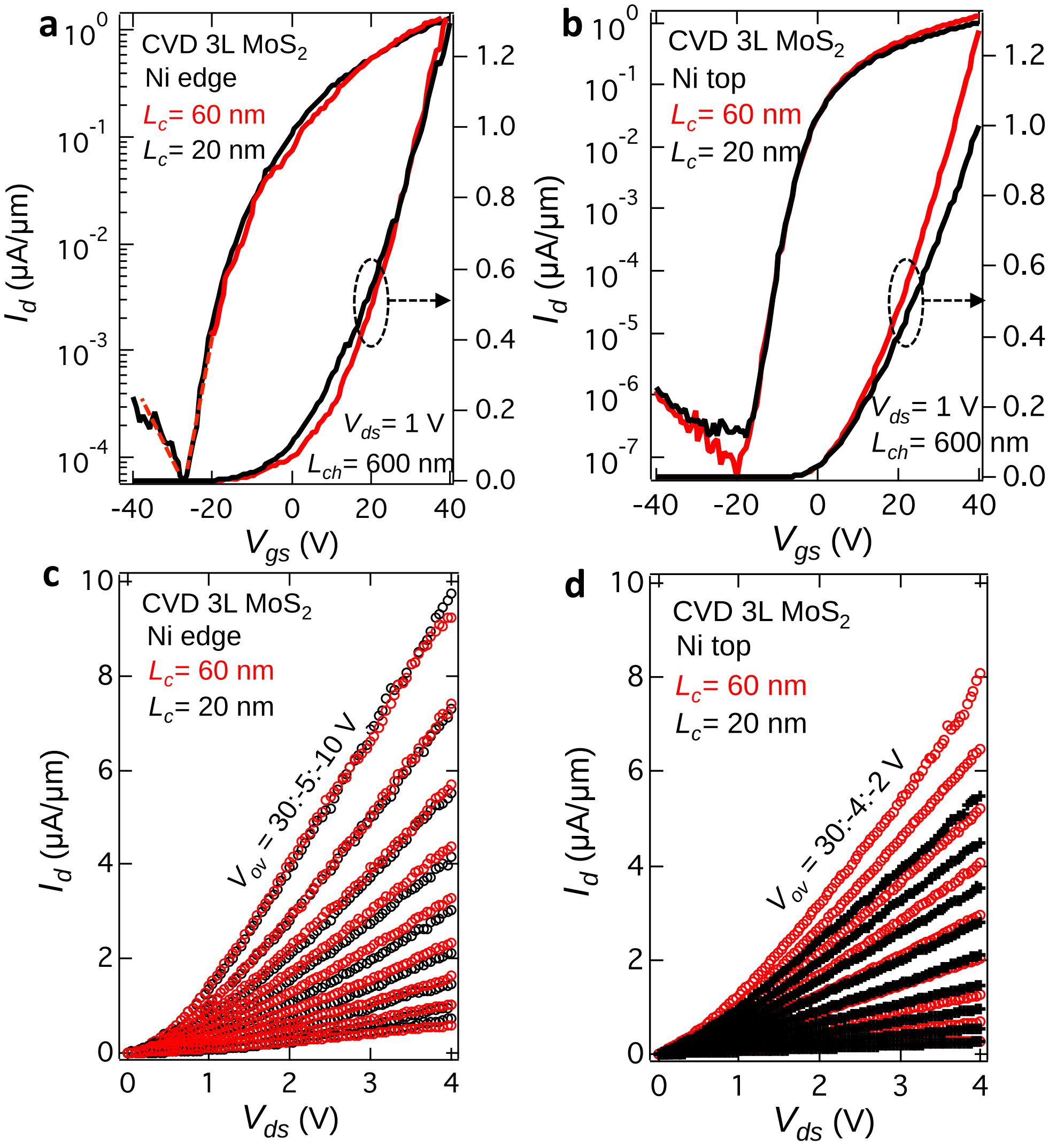

**Supplementary Figure 6. Comparison of edge- and top-contacted devices using Ni as contact metal.** $I_d$-$V_{gs}$ curves for Ni edge-contacted (**a**) and top-contacted (**b**) $MoS_2$ FETs. The red curves in (**a**) are shifted in order to have the same $V_{th}$ and have a fair comparison. Output curves for the Ni edge-contacted (**c**) and top-contacted (**d**) $MoS_2$ FETs. The decrease in $L_c$ leads to performance degradation in top-contacted devices but has little impact for edge-contacted devices.

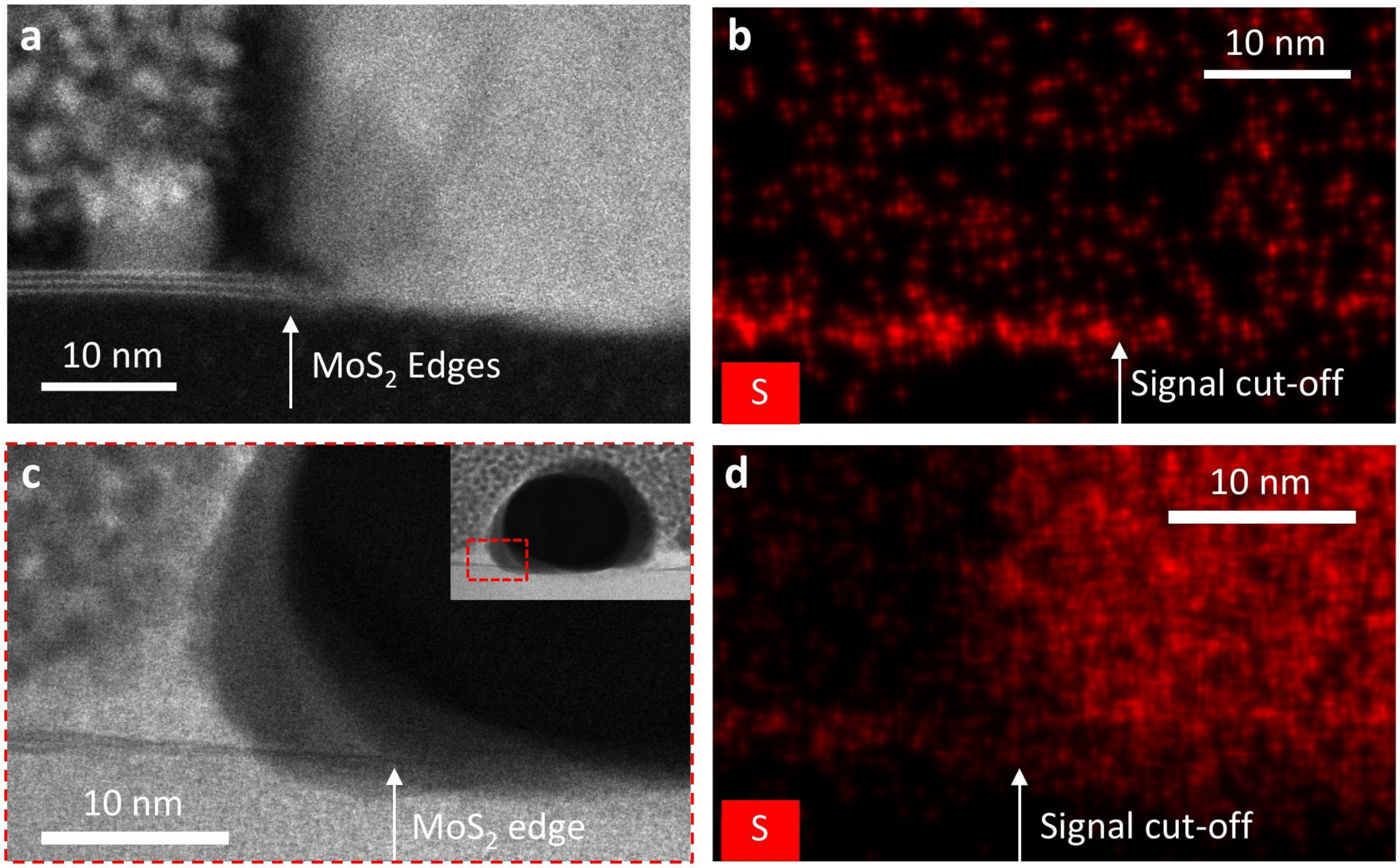

**Supplementary Figure 7. EDS mapping sulfur at the edge interface.** Cross-sectional STEM images of the edge contact to **a**, trilayer and **c**, monolayer $MoS_2$. **b** and **d** are EDS images of sulfur signal in the edge area of **a** and **c**, respectively. The discontinued trace within the contact metal suggests some sulfur residue. The sulfur signal is weaker in the monolayer $MoS_2$, compared to the signal in the trilayer $MoS_2$.

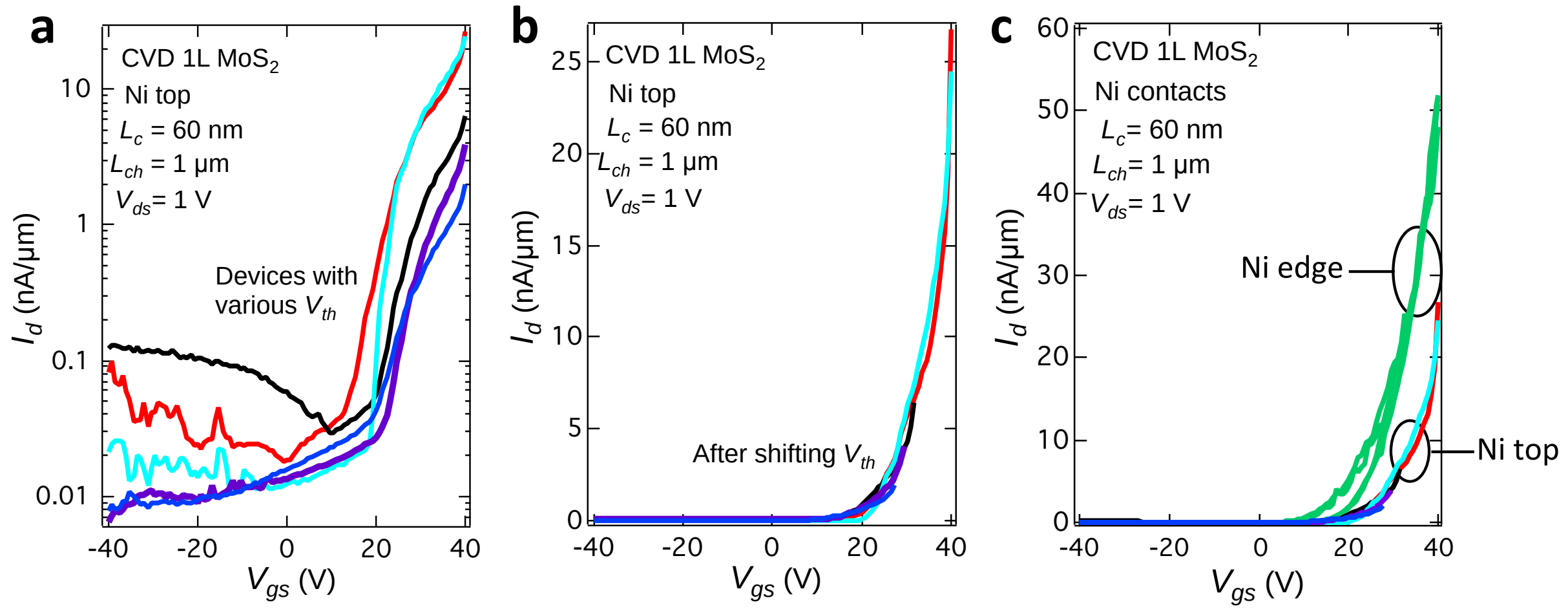

**Supplementary Figure 8. Comparison of top and edge contacts using Ni to monolayer $MoS_2$. a**, Subthreshold curves and **b**, transfer curves of the edge contacts after shitting the $V_{th}$ in (a). **c**. Various Ni edge contacted devices (green) comparing with Ni top contacted devices showing similar current level with different $V_{th}$. The small $I_d$ for both the top- and edge-contacted devices indicates that the quality of the CVD films (grown on $SiO_2$) dominates the performance of the devices.

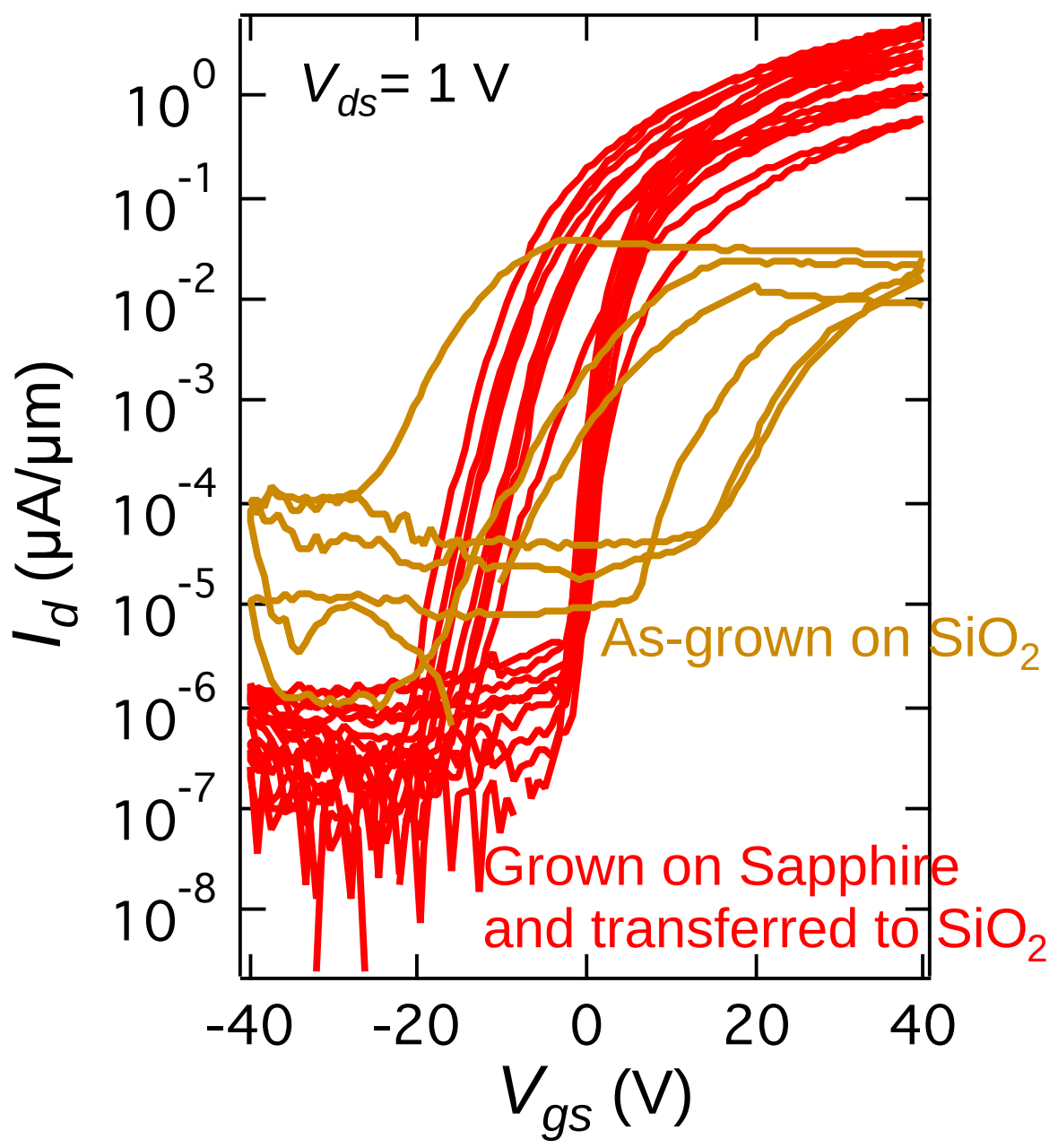

**Supplementary Figure 9. Comparison of $I_d$-$V_{gs}$ curves for transistors using as-grown $MoS_2$ versus transferred $MoS_2$.** The thickness of the CVD films ranges from 1 to 2 layers. The channel length of all transistors is from 1 to 3 μm. The transistors built on transferred $MoS_2$ have a larger $I_d$ and on/off ratio, and smaller hysteresis. The transfer process leads to fewer traps between the $MoS_2$ and the substrate, improving the overall device performance.

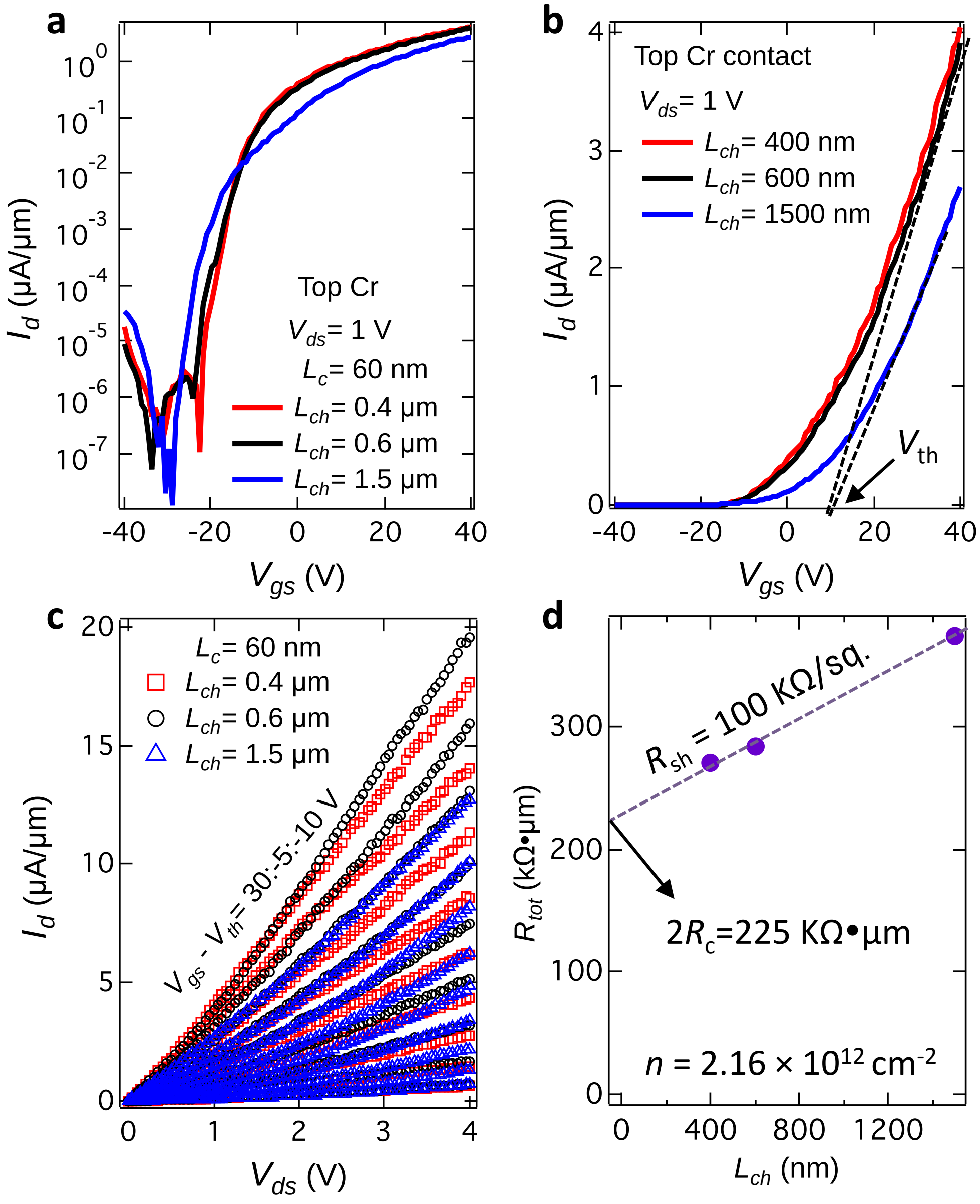

**Supplementary Figure 10. Top Cr contacts to 3L $MoS_2$ grown by CVD and contact resistance extraction**. **a**, $I_d$-$V_{gs}$ curves and **b**, $I_d$-$V_{ds}$ curves for top Cr contacts with transfer length method (TLM) structure. **c**, Extraction of the contact resistance ($R_c$ = 110 kΩ•µm) for Cr top contacts. The sheet resistance of the channel is around 100 kΩ/sq at the overdrive voltage of 30 V. The high $R_c$ and $R_{sh}$ can be attributed to 1) the high density of traps formed between the SiO2 and MoS2 during the high temperature growth of CVD, and 2) the low carrier density of n = 2.16 × $10^{12}$ cm$^{-2}$ ($V_{ov}$=30 V over 300 nm $SiO_2$), compared to other reports in Supplementary Table 3.

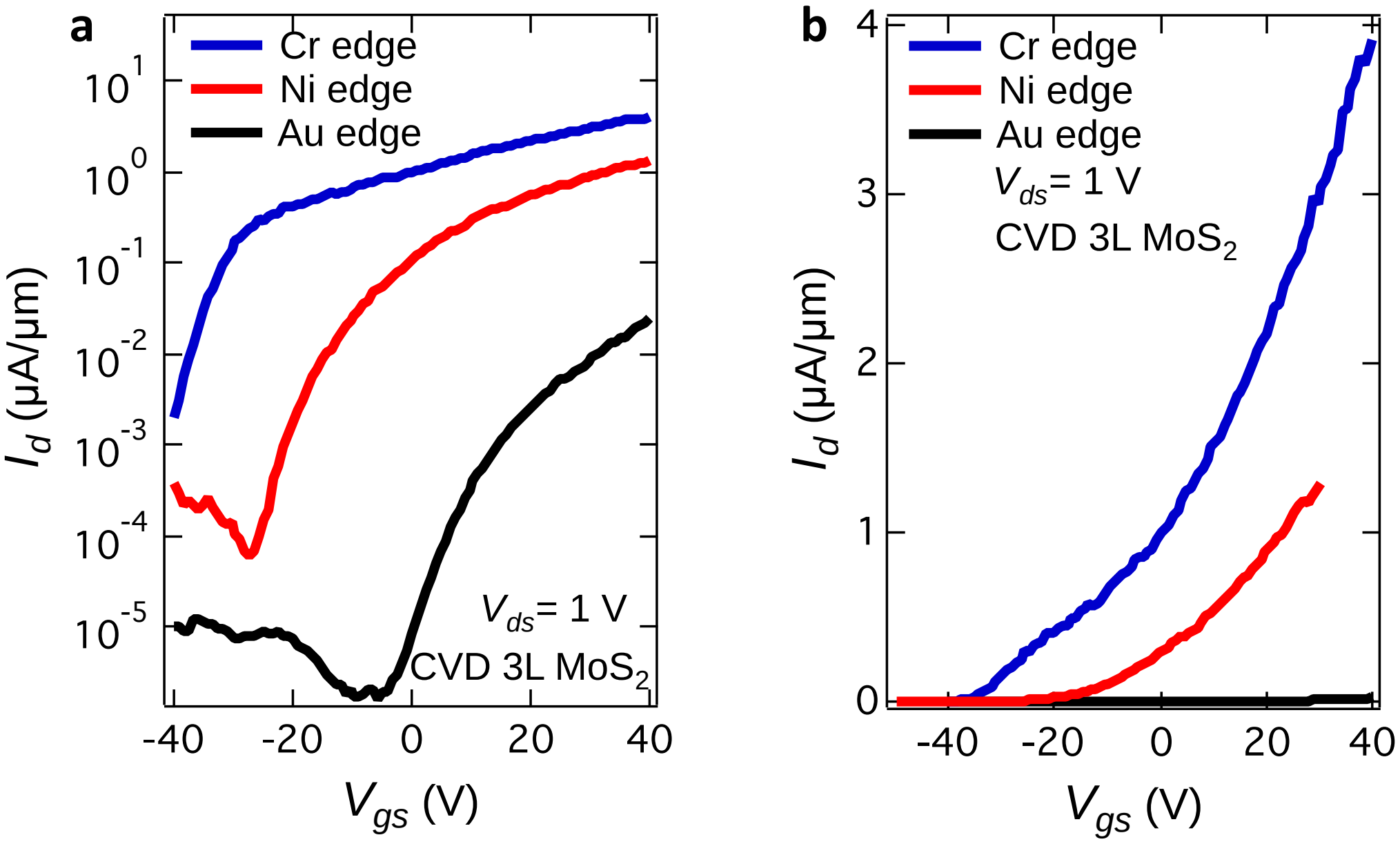

**Supplementary Figure 11. Comparison of edge contacts using Cr, Ni, and Au**. **a**, Subthreshold curves and **b**, transfer curves of the edge contacts. Note: Curves were shifted in b to have the same threshold voltage for on-state comparison. The $L_{ch}$ and $L_c$ for these devices using different metals are 600 nm and 60 μm, respectively.

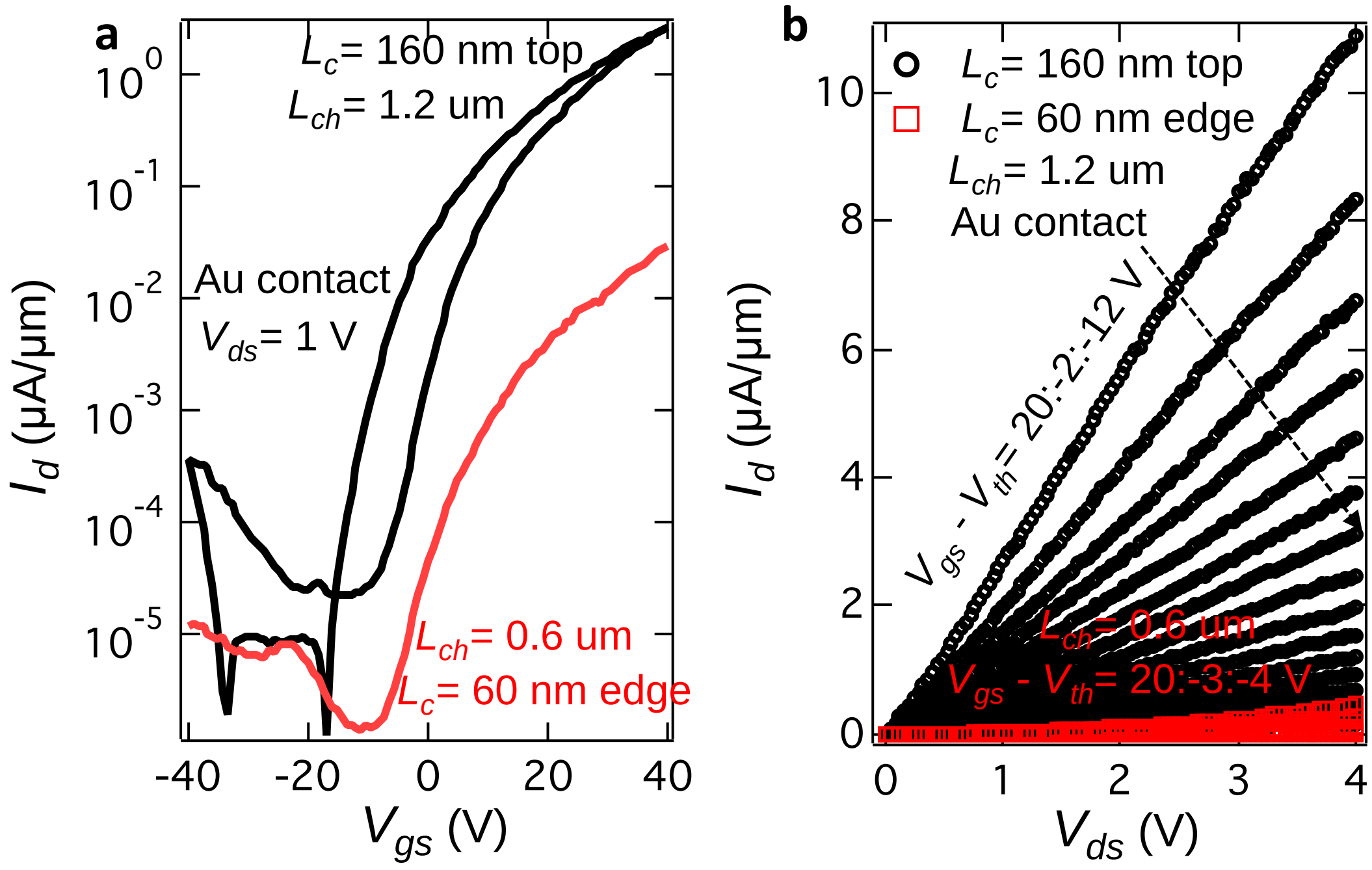

**Supplementary Figure 12. Comparison of Au top and edge contacts to 3L $MoS_2$ grown by CVD.** **a**, Subthreshold curves and **b**, output curves of the Au contacts. Even though the devices are with different channel length and contact length, their dramatic difference indicates the huge contact resistance of Au edge contacts.

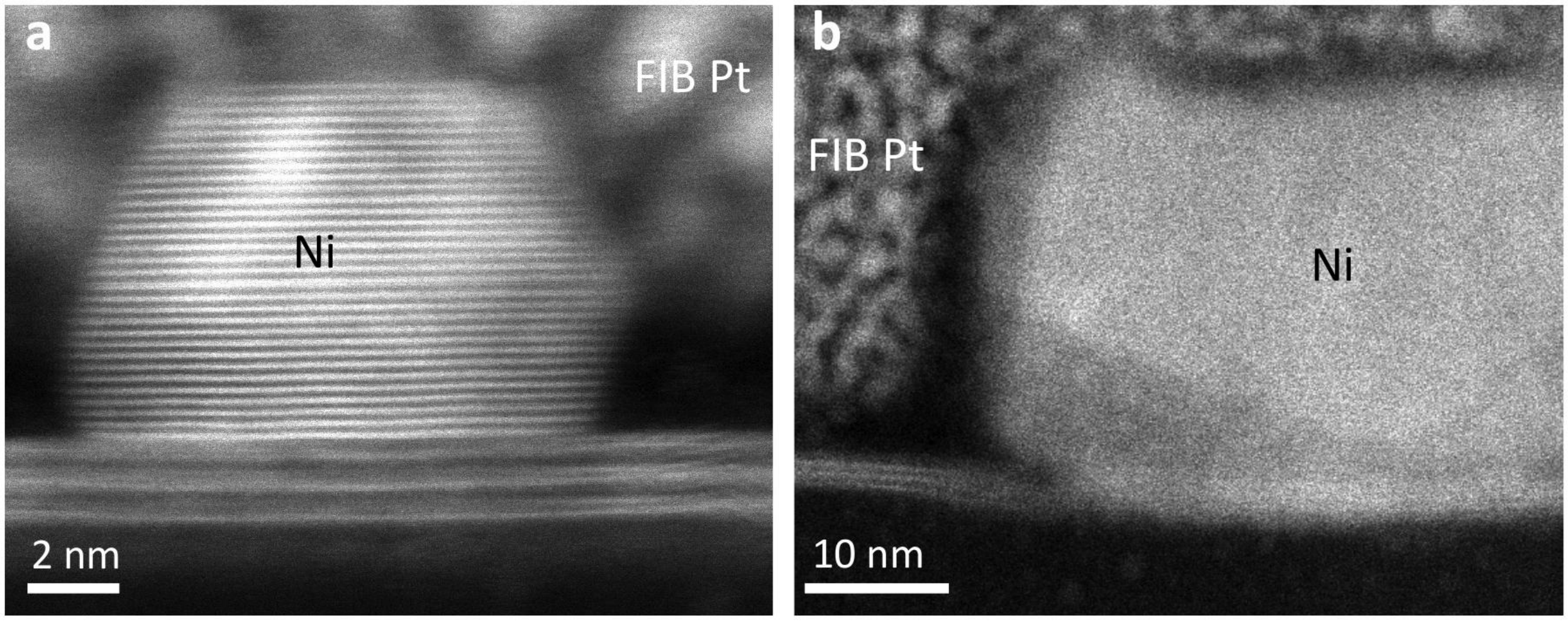

**Supplementary Figure 13. Different profile of Ni top and edge contact under cross-sectional STEM imaging**. **a**, 10 nm top contact on a trilayer $MoS_2$ flake and **b**, 40 nm edge contacts to a trilayer $MoS_2$ flake.